# TDMP: Reliable Target Driven and Mobility Prediction based Routing Protocol in Complex Vehicular Ad-hoc Network

M.Ye[a,*], L.Guan[a], M.Quddus[b]

[a] Department of Computer Science, Loughborough University, Loughborough LE11 3TU, United Kingdom

[b] School of Architecture, Building and Civil Engineering, Loughborough University, Loughborough LE11 3TU, United Kingdom

## Abstract

Vehicle-to-everything (V2X) communication in vehicular ad hoc network (VANET) has emerged as a crucial component in advanced Intelligent Transport System (ITS) for information transmission and vehicular communication. One of the vital research challenges in VANET is the design and implementation of novel network routing protocols which bring reliable end-to-end connectivity and efficient packet transmission to V2X communication. The organically changing nature of road traffic vehicles poses a significant threat to VANET with respect to the accuracy and reliability of packets delivery. Therefore, position-based routing protocols tend to be the predominant method in VANET as they overcome rapid changes in vehicle movements effectively. However, existing routing protocols have some limitations such as (i) inaccurate in high dynamic network topology, (ii) defective link-state estimation (iii) poor movement prediction in heterogeneous road layouts. Therefore, a novel target-driven and mobility prediction (TDMP) based routing protocol is developed in this paper for high-speed mobility and dynamic topology of vehicles, fluctuant traffic flow and diverse road layouts in VANET. To implement an effective routing protocol, TDMP primarily involves the destination target of a driver for the mobility prediction and Received Signal Strength Indicator (RSSI) for the inter-vehicular link-status estimation. Compared to existing geographic routing protocols which mainly greedily forward the packet to the next-hop based on its current position and partial road layout, the proposed TDMP is able to enhance the packet transmission with the consideration of the estimation of inter-vehicular link status, and the prediction of vehicle positions dynamically in fluctuant mobility and global road layout. Based on the extensive simulations carried out on operational road environments with varying configurations and complexity, the experimental results show better performance in terms of improving packet delivery ratio by 21-57%, reducing end-to-end delay by 13-47% and average hops count by 17-48% in comparison with several typical position-based routing protocols, such as GPSR, GyTAR and PGRP.

## 1. Introduction

Nowadays, a tremendous evolution of advanced technologies and sophisticated solutions applied to Intelligent transport systems (ITS) has been observed. For instance, Internet of Vehicle (IoV) allowing both appealing infotainment systems and traffic management applications which require internet access is a core component of future ITS. For vehicular interactions, a short-range communication technology incorporating GPS-equipped vehicles and stationary roadside units (RSUs) has been widely used, which is defined as Vehicular Ad-hoc Network (VANET) [1]. VANET exploit Inter-Vehicle Communication (IVC) protocols to become the key part of Cooperative-ITS [2]. Also, with the advances in wireless communication technology, the concept of a networked car has received immense attention all over the world. This kind of importance has been recognised by the major government organizations, industrial manufacturers and academic research.

In VANET, each vehicle acting as the network node communicates with another vehicle and constitutes a large ad-hoc network. Considering a huge number of vehicles (expected up to 2 billion on the world's road by 2035), the market and benefit of VANET would increase exponentially in the future. For example, VANET can be utilised for real-time traffic data collection for both safety and non-safety applications, advertisement propagation, advanced navigation calibration, location-based services, parking information sharing, infotainment applications and internet access. Therefore, V2X communication including vehicle-to-vehicle (V2V) and vehicle-to-infrastructure (V2I), an infrastructure-free mechanism, will be in great demand for reliable and efficient information transmission (e.g., vehicular kinematic information, traffic conditions, sales news and interactive messages) sooner or later.

The pivotal requirement for the achievement of VANET applications is the availability of one robust routing protocol for messages dissemination. In order to enable geographically separated vehicles to link together, VANET adopts multi-hop wireless communication by relying on intermediate vehicles for data transmission to extend the coverage of vehicular communications and internet-based services [3]. For more reliable and sustainable connectivity, automotive manufacturers employ cellular network for inter-vehicle internet access. However, in the high-density traffic area, and with respect to the explosive growth of mobile data traffic, the centric cellular networks cannot afford the high communication overhead. It is measured that the current mobile data demand will increase over 10 times and the monthly mobile data traffic will exceed 77 exabytes by 2022 [4]. Hence, a hybrid network of VANET and cellular network can be deployed to both effectively support VANET users with low-cost internet-based services and greatly mitigate the cellular network overload [5].

The high-speed mobility of road traffic vehicles and heterogeneous road layouts cause rapid changes in vehicles density and intermittent inter-vehicular communications. Moreover, the existence of obstacles, such as large vehicles and building, can hugely influence the radio signal and disrupt inter-vehicular data transmission, even when vehicles are within the communication range [6]. For the purpose of mitigating the influence of highly dynamic topologies and guaranteeing inter-vehicular connections, one of the most challenging tasks to address unique characteristics in VANET is the design and implementation of communication routing protocols. Most of the existing routing protocols hardly take both of aforementioned two factors into consideration. VANET is slightly different from Mobile Ad-hoc

Network(MANET) by its characteristics, requirements, architecture, challenges and applications [7]. Therefore, conventional routing protocols used in MANET cannot be used directly in the field of VANET because of unwarranted performance. Massive works have been done to solve routing issues in VANET, such as position-based routing (PBR) protocols, cluster-based routing protocols and regional-multicast routing protocols.

Moreover, spatio-temporal geographical positions of the vehicles can be easily accessed by their GPS devices nowadays, and the mobility of vehicles are supposed to follow the fixed road segment in a particular pattern. In the implementation of routing protocol, destination vehicle's position can be provided to the source vehicle by some location services such as Grid Location Service (GLS), Hierarchical Location Service (HLS), Reactive Location Service (RLS) and Semi-Flooding Location Service (SFLS) [8]. With the usage of advanced devices, it is possible to detect and predict vehicles' mobility pattern, which effectively supports the packet forwarding with higher accuracy and universality. At the same time, Received Signal Strength Indicator (RSSI) based techniques are low-cost methods without any specialized hardware, which is from the idea that receiver can estimate link quality with the sender by RSSI values using theoretical radio propagation models [9]. RSSI-based methods are ideal to measure the stability of V2V connections, which can practically improve the reliability of packet forwarding. Meanwhile, most of our journeys are triggered by a target-driven route with increasingly common usage of navigation services. Popular online services, such as Google Maps, Waze and GAODE Map, can provide online dynamic navigation guidance for users based on real-time traffic status information collected from vehicles or mobile devices [10]. With the support of the in-vehicular network, inter-vehicular communication accessories can easily obtain the target information of drivers, such as the interest of places and destination. These types of mobility-related information are potentially beneficial for routing and forwarding packets in VANET.

Based on the aforementioned issues, we develop a novel target-driven mobility-prediction based routing protocol(TDMP) in this paper. TDMP considers RSSI values, predictable mobilities of vehicles and the target of the receiving-end vehicle while routing and forwarding packets.

The main contributions of this paper are listed as follows:

(1) Combining both enhanced forward strategy and recovery strategy with new mechanism involved in messages transmission can greatly decrease the packet loss and delay, and ameliorate network overhead.
(2) Originally introducing the Original/Destination Demand Matrix to represent both navigation system of vehicles and target information of drivers.
(3) In order to overcome the inherent constraints of local road layouts, the target-driven mechanism can select a better relay to forward messages on a global scale.
(4) Involving the RSSI to measure the vehicles' link status beforehand is able to efficiently improve the reliability of packet routing and forwarding.
(5) Analysing and discussing to what extent different factors may influence the neighbour selection.
(6) Validating the effectiveness and feasibility of the developed protocol by adopting a unified simulation platform (Veins) with different road scenarios and implement them in the real-world situation. In addition, a series of comparisons with existing routing protocols in the related work has been carried out to prove the improvement.

The rest of the paper is organised as follows: Section 2 gives a brief description of related works. Section 3 describes the detailed TDMP routing protocol. The details of the analysis and simulation-based performance evaluation are shown in section 4. In section 5, we provide the conclusion and future prospects.

## 2. Background and related work

Many types of routing protocols have been proposed for VANETs, as surveyed in [3][11][12][13][14][15]. Since our work is closely related to the position-based process, we mainly review and discuss works in this category. Particularly, the routing protocols that are relevant to our approach are discussed at a greater depth.

### 2.1 Classification of routing protocols in VANET

Broadly speaking, existing VANET routing protocols can be systematically classified into two main categories: (i) V2V and (ii) V2I respectively. There are mainly four types of V2V routing protocols: topology-based routing protocol, position-based routing protocol, cluster-based routing association, and regional multicast routing protocol as shown in Fig.1.

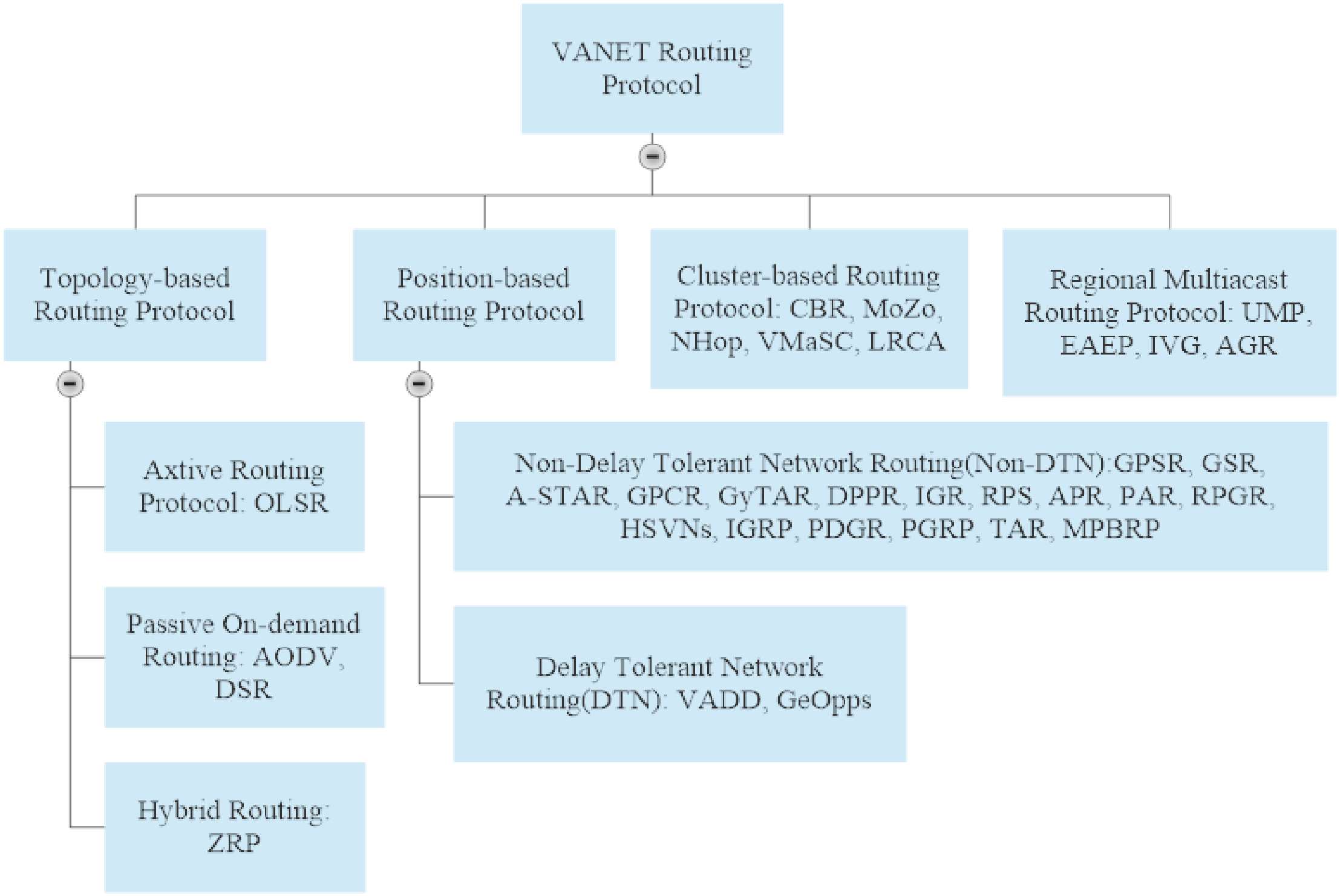

Fig.1. Taxonomy of VANET routing protocols

Topological routing forward data through existing links in the network, which includes active routing, passive on-demand routing and hybrid routing driven by routing table. Unlike other networks, vehicles' high mobility and frequent change of communication links between vehicles make the traditional topology-based routing protocols, such as OLSR (active routing) [16], AODV (passive on-demand routing) [17][18], DSR (passive on-demand routing) [19] and ZRP (hybrid routing) [20], fail in VANET because they flood the packets with extensive pathfinding and maintain control messages, which caused increased routing load and network security problems. The link found in passive routing is likely to be disconnected soon, so this type of routing is not suitable for the vehicle-borne network.

To overcome the disadvantages of table-driven topological routing, an alternative geographical location-based routing paradigm or position-based routing has been introduced [3][15][21][22][23][24]. In PBR, vehicles need to collect the position information of themselves and their neighbours. PBR can be divided into non-delay tolerant network (Non-DTN) routing and delay-tolerant network (DTN) routing. The detailed introduction is presented in section 2.2 below. Existing studies have confirmed that this paradigm, PBR, outperforms traditional topology-based routing in both urban and highway VANET's scenarios [3].

Clustering routing arranges vehicles into clusters and only need the cluster heads to maintain neighbouring information, which is generally more suitable for the network with clustering topology. One vehicle in the cluster is selected as the head which is responsible for intra-cluster and inter-cluster communications, while other nodes can only communicate directly with nodes in the same cluster. The formation of the cluster and the selection of the cluster head is very important in this mechanism, which is mainly influenced by the network types. Some typical clustering routing, such as CBR [25], MoZo [26], VMaSC [27] and LRCA [28], have good performance in small networks and some urban areas with high-density traffic flow. However, clustering routing shows really poor performance in the suburban area with an insufficient number of vehicles.

Regional multicast routings rely on large message dissemination in a region and hence may cause a high communication overhead and message congestion on the network. Typical routings include EAEP [29], IVG [30] and AGR [31]. Another serious drawback of such protocols is network partitioning and the presence of harmful neighbour nodes, which can hinder the proper forwarding of messages. However, this mechanism can guarantee the receiver to get the information effectively and accurately.

### 2.2 Position-based routing (PBR) approaches

Position-based routing is a connectionless routing approach in which the establishment of communication process is not needed before data transmission, and data packets are routed independently [3]. Many schemes in PBR have explored global positioning, relative positioning and surrounding region based attributes to identify their road segments and junctions based on vehicle position [32]. In PBR, each vehicle necessitates updating the kinematic information of itself and its neighbouring vehicles for future tracking and analysis. And such routing mechanism does not need to maintain the routing tables. Accordingly, PBR is considered a more promising routing approach for dynamic environments, since it provides scalability and robustness against frequent topology changes [15].

PBR can also be classified into Non-DTN and DTN routing protocols. The aim of the Non-DTN is to impart data packets to the destination node as soon as possible by exchanging road information rapidly, which is greatly used in the effectively populated VANETs, including GPSR [33], GSR [34], A-STAR [35], GPCR [36], GyTAR [37], DPPR [38], IGR [39], RPS [40], APR [41], PAR [42], RPGR [43], HSVNs [44], IGRP [45], PDGR [46], PGRP [47] and MPBRP [48]. Different from non-DTN, DTN with carry and forward mechanism will not move the data packet until establishing the stable node connections, which includes VADD [49] and GeOpps [50]. In VANETs, vehicular density has an important impact on inter-vehicle communication links stability. In order to eliminate traditional PBR limitations, new metrics have been introduced by recent routing protocols, which involve network and traffic status in routing decisions. Integrating PBR with traffic awareness results in traffic-aware routing (TAR) protocols, which consider variable traffic conditions and diverse road layouts. TDMP routing protocol is inspired by the following position-based routing protocols, also it has been partially compared with them in Section 4.

Greedy Perimeter Stateless Routing (GPSR) is the fundamental position-based routing protocol proposed for ad hoc networks, whose basic mechanism is the greedy algorithm, which enable messages to reach the destination as soon as possible in the dense network. As GPSR forwards packets only greedily based on vehicles' position without consideration of traffic and network status, packets might be forwarded through roads with low vehicular density or high level of network disconnections, which will greatly increase packet loss rate and transmission delay [33].

Greedy Perimeter Coordinator Routing (GPCR) scheme improves the trustworthiness of GPSR in VANET [36]. Basically, GPCR works like GPSR, but one change is that GPCR chooses a relay node by analysing the road information. GPCR considers junction-based routing and its position together, rather than only selects a single one. In GPCR, vehicles at the junction forward packet by analysing traffic density on the adjacent node and connectivity of that node to the destination. If traffic density is low and the connection is obviously weak between nodes and destination, latency and transmission delay may increase due to local maximum problem. GPCR considers centred vehicles at the intersection as a special vehicle called the coordinator to relay the packets and solve the obstacle problems. And packets in GPCR are directly delivered between the junctions without the usage of road maps, which may fail in the selection of the best path.

An improved Greedy Traffic-Aware Routing (GyTAR) protocol is applicable to the urban environment, which is an intersection-based geographical routing protocol [37]. GyTAR takes into account unique features of the vehicular environments that include high dynamic vehicular traffic, road traffic density, and road topology, for both car-to-car communication service and value-added infrastructure-based ITS services [51]. The data packages are forwarded greedily and routed to their destination through the intersection. Besides, the dynamic selection of the intersection relies on its curvilinear distance towards the destination node and the traffic density between the current intersection and the candidate intersection. However, the lack of global information would cause the wrong selection of the real optimal intersection, and then increase the transmission delay and the packet drop.

Predictive Directional Greedy Routing (PDGR) selects the next-hop neighbour by taking into account the position, direction and the speed of each neighbour, and using a directional greedy mechanism [46]. The prediction process of PDGR is represented as the weight calculation for each next-hop node by using their position and direction information. NS2 simulator has been used to test and evaluate PDGR. Packet delivery ratio, end-to-end delay, average hops, send rates and number of nodes have been selected to test the PDGR. However, PDGR is only simulated for an open environment without considering the urban area.

Predictive Geographic Routing (PGRP) is one latest routing protocol with highlighting vehicle connectivity problem, which improves the routing performance by considering the mobility constraints and predictable natures of vehicles [47]. The MOVE platform combining SUMO and NS-2 has been used to test the PGRP. PGRP can be used in both the grid-based environment and the highway environment. However, PGRP is lack of using the acceleration of each vehicle and the driver's target to make a better decision in a real urban environment.

Recently, some researchers also present the FoG-oriented framework for PBR scheme in VANET [32]. They utilise linear programming, genetic algorithms and regression-based schemes for considering connectivity aware PBR schemes in the urban environment. In their work, road junctions are selected for path selection and parked vehicles near junctions are chosen for packet transmission. This type of framework aims to support better packet delivery ratio, end-to-end delay, transmission time, and communication cost.

## 2.3 Discussion and knowledge gap

For reliable and timely message transmissions, PBR protocols involve the position of a moving vehicle to assist in dynamic path discovery. Most of the position-based routing protocols suppose that the velocity of vehicles is static during the transmission of beacon messages without taking into account highly variable speed and direction changing in VANET. For example, within the transmission range of the source vehicle, if the closest vehicle to the destination runs in a different direction and out of the communication range later, it would cause the high end-to-end delay and decease the packet delivery ratio seriously due to the lost connection of chosen vehicle. Besides, acceleration and deceleration process cannot be ignored in the prediction of vehicle mobility.

This paper develops a novel position-based routing protocol known as Target Driven and Mobility Prediction based routing protocol (TDMP) with considering the obstacle ratios in the urban scenario, acceleration and drivers' intention, which is tested with a novel unified simulation platform. The results show great improvements compared with some existing routing protocols stated above. GPSR is not considered to be efficient because the neighbouring table is not updated and it may cause the highest delay because there is no updated information on neighbouring vehicles. The first neighbour obviously changes its position and new vehicles take its previous position. It may be a case that a new vehicle which moves to the source node but the source node has not updated its information. But at the same time, GPSR is also good because it only considers single-hop radio neighbours and dynamically decides the packet forwarding. However, it will become vulnerable in case of density variations.

On the contrary, GeOpps is not affected by the higher density of vehicles on road. If there are many vehicles on the road and the source node wants to send messages to any other vehicle, then it is not difficult to select the reliable neighbour node. But lower vehicles densities may reduce the chances of connectivity and result in poorer packet delivery ratios. In these protocols, only GPCR and GyTAR are obstacle-aware protocols and other protocols are only street-aware protocols. GPCR and GyTAR are good for city environment due to obstacles awareness. PGRP involves the prediction of vehicles' position according to vehicle position information gathered from GPS and beacon message so as to solve real-time V2V communication both in highway and urban scenarios [47]. But with the development of GPS-based navigation and route guidance technology, the destination information of driver can be easily obtained, which can be potentially applied in the selection of the best neighbouring node. The proposed TDMP routing protocol can significantly counteract the disadvantages and problems in the mentioned position-based routing protocols, and then achieve better performances.

## 3. TDMP: Proposed Target-Driven and Mobility Prediction based routing protocol

### 3.1 Overview of TDMP

VANET supports a series of applications and services to a certain extent utilizing the IEEE 802.11p protocol, such as traffic warnings of emergencies and dynamic route planning in the congestion zone. Emergency messages can be triggered by the event-driven mechanism. Event-driven messages are sent when necessary such as a warning of the sudden braking or a vehicle is involved in a crash in a critical situation. Particularly, the information on the number of casualties should be updated promptly and accurately with the upcoming ambulance to keep them safe within the golden one hour because casualties have a much slim possibility to survive if they cannot receive definitive care within one hour [52]. As shown in Fig.2, the emergency message is triggered by a serious crash event and it is updated periodically and transferred to the nearby ambulance service centre by VANET. Within this case, real-time information, such as the crash location, voice messages and potential videos, are sent from the crash spot to the upcoming ambulance. This end-to-end connection in VANET can improve the efficiency and accuracy of information transmission in the complex network environment.

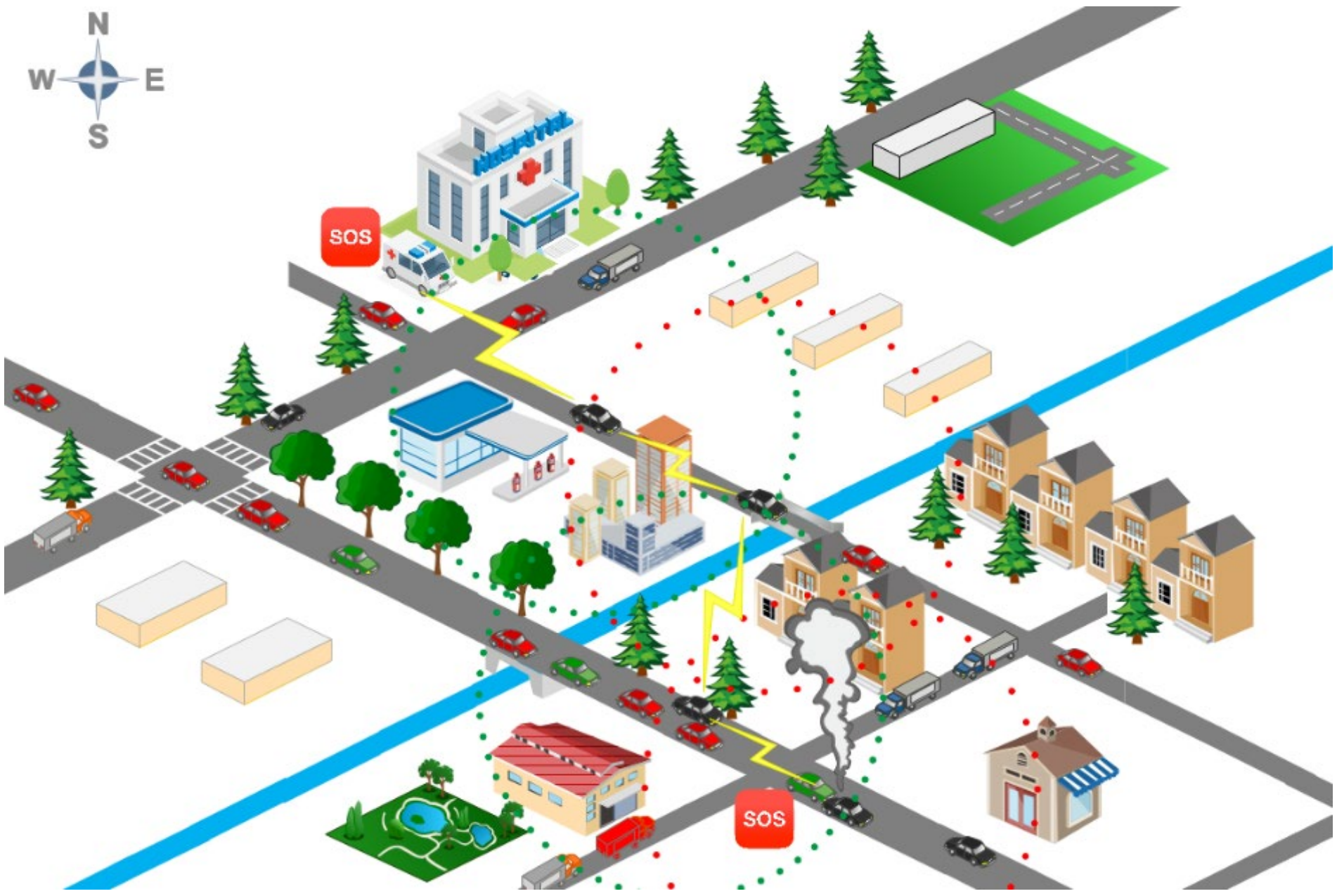

Fig.2. Event-driven accident alert in VANET

TDMP assumes that each vehicle in the network has the knowledge of its own position, velocity and acceleration by onboard GPS, the kinematic information of its neighbour by information exchange and its destination on a digital map by using the equipped location services. The knowledge about the location of a destination node in GPSR is assumed to be available for the source node [50]. In GPSR simulation, all mobile nodes can be provided with the location information without any cost [30]. Thereby, TDMP also presumes the availability of a destination's location where the source node forwards and routes the packet. In a real-world scenario, the destination's location can be informed timely by location-based services using city-scale wireless sensor networks. In Fig.2, kinematic information of the ambulance can be shared periodically with nearby vehicles. In our simulation, the Original/Destination (O/D) Matrix can be a reasonable solution to achieve the destination information of each vehicle. Meanwhile, each vehicle maintains a data table where its coordinate, velocity, acceleration and the location of its destination are recorded. This table is established and updated by periodic exchange of beacon messages among all vehicles running on the road. Also, each vehicle can measure the received signal strength indicator for the estimation of inter-vehicular connectivity, which can be reached by IEEE 802.11 package in the simulation. Vehicles are capable to communicate with each other within 300 meters.

### 3.2 TDMP algorithm and its description

This section explains the developed TDMP routing protocol for V2V multi-hop communications in VANET environment. TDMP makes its routing decision by both estimating the inter-vehicle link status and predicting the spatio-temporal movements of the vehicles, which is ideal and practical both in low-density and high-density traffic scenarios. Through estimating the inter-vehicle link status, TDMP can perceive the nearby building blocks and a large number of vehicles which would influence the packet transmission. By predicting the movements of neighbouring vehicles and calculating the weighted score, TDMP would select the best neighbour so as to forward the information to the destination. To obtain the mobility of neighbouring vehicles for packet routing,  a novel method is developed by combining the predicted position, potential direction and target of vehicle. Moreover, the involvement of O/D matrix in the simulation is a creative way to represent the vehicle's target. O/D matrix can also indicate traffic demand between specified Traffic Analysis Zone(TAZ). Based on the hypothesis in Sec.3.1, each vehicle equipped with advanced GPS-based navigation and route guidance systems can retain its target destination, which is the core part of the TDMP routing protocol.

In terms of position-based routing protocols, two main issues should be resolved, the forwarding mechanism and the recovery mechanism [3]. In order to address the local maximisation problem, two special forwarding strategies are used: the predictive greedy forwarding algorithm and the predictive perimeter forwarding algorithm. The greedy forwarding algorithm employed in TPDM differs from the conventional forwarding strategies in some respects. The vehicles utilising the greedy forwarding strategy would broadcast beacon messages and then select the closest neighbour towards the destination based on the geometric heuristics [33]. As shown in Fig.3, the source node $A$ would select the adjacent node which is the closest to the destination node $Z$, and then forward the package to the selected node $B$. For example, the packet can be sent to the destination by the route $A \rightarrow B \rightarrow D \rightarrow E \rightarrow Z$ in Fig.3 if the topology is assumed to be stationary in the transmission process. This greedy forwarding strategy is effective to deal with the static nodes. However, given the high-speed vehicle mobility in a VANET environment, conventional forwarding strategies are inappropriate herein.

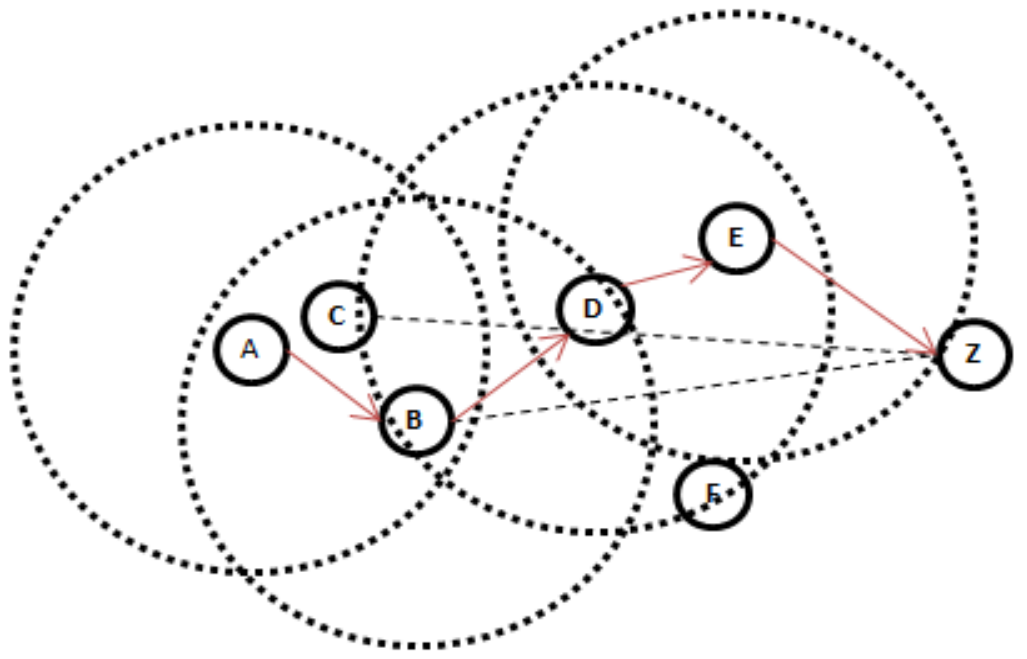

Fig.3.The Greedy Forwarding Strategy

Moreover, there is an issue with the problem of the local maximum in the greedy forwarding algorithm because of limited global information.  For instance, if sender $A$ detects node $B$ and $C$ within the transmission range as shown in Fig.4, but none of them is closer to the destination node $Z$  as  $d_{AZ} < d_{CZ} < d_{BZ}$. This implies that the packet traps to a local maximum. Herein, node B and C cannot become the potential candidates to forward the packet by the greedy forwarding strategy. However, the enhanced perimeter forwarding algorithm in TDMP can resolve this issue. Two recovery methods are dominant: Relative Neighbourhood Graph (RNG) and Right-Hand Rule [33]. To overcome the problem with the local maximum, RNG and Right-Hand Rule are used to check the connectivity to the destination. In Fig.4, node $C$ would be selected to relay the packet. Eventually, the packet can be sent to the destination by the route $A \rightarrow C \rightarrow D \rightarrow E \rightarrow Z$ . However, the high mobility of VANET is also ignored in this strategy.

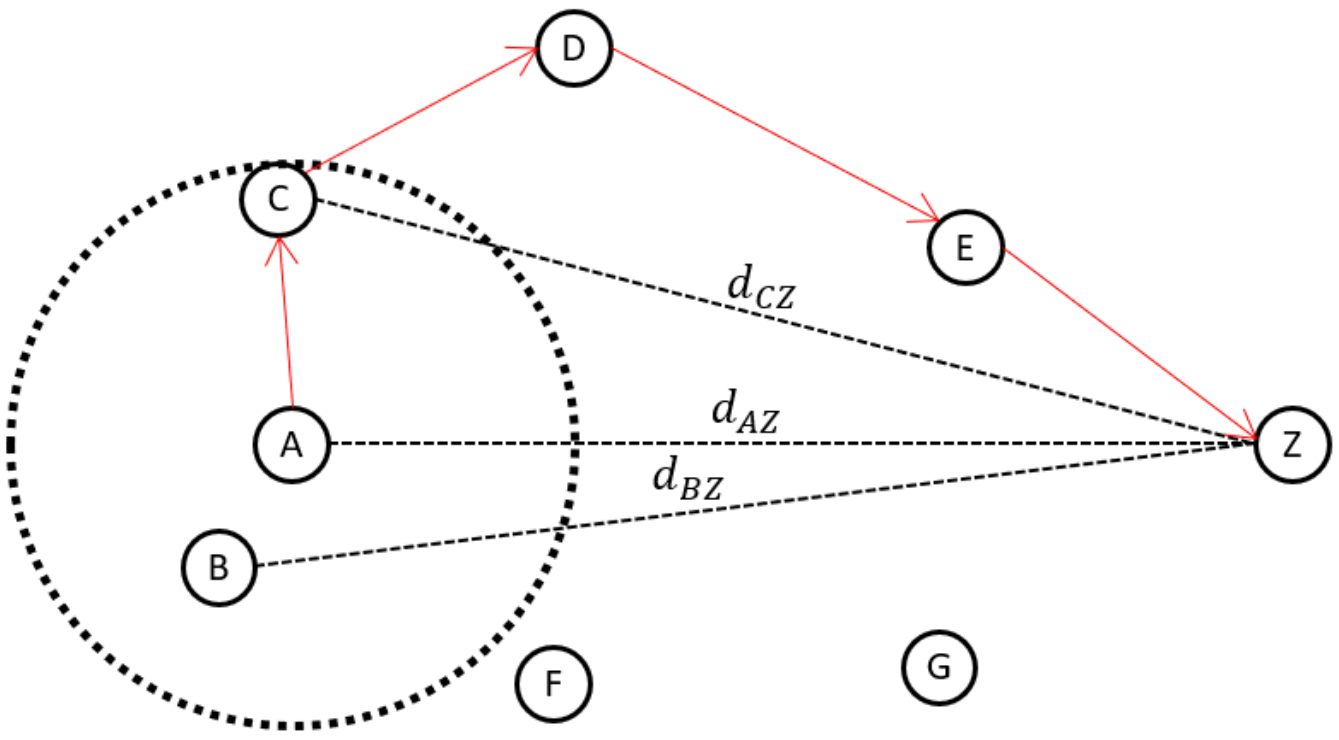

Fig.4.The Perimeter Forwarding Strategy

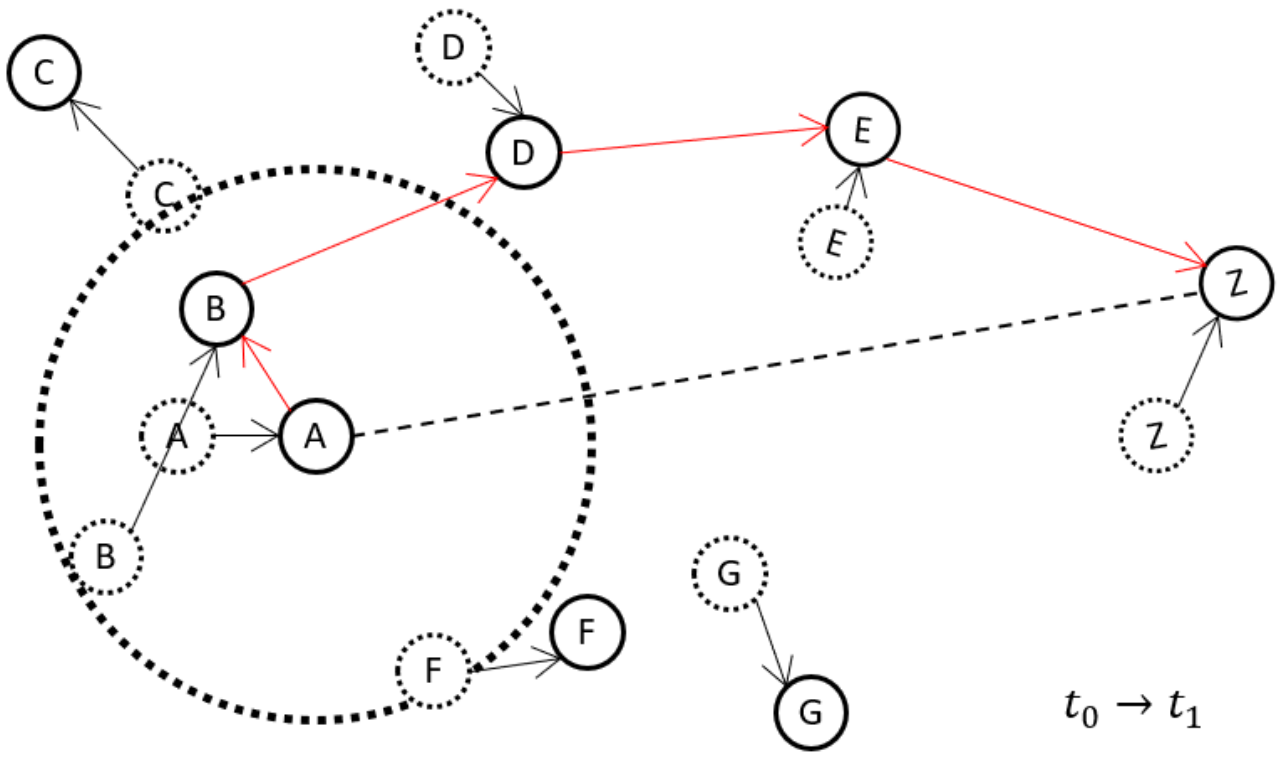

Fig.5.The Predictive Perimeter Forwarding Strategy

Therefore, the prediction of the future positions of the vehicle is important in the forwarding strategy, which can support both the greedy forwarding strategy and the perimeter forwarding strategy for the selection of the best neighbour. In the TDMP simulation, it takes one second to trigger the source node to broadcast 'hello' messages and receive the responses from the neighbours. For example, assuming that node *C* in Fig.4 moves at a high speed to a certain direction at time $t_0$, shown in Fig.5, chosen node *C* would change the position and be out of bounds at time $t_1$. The packet relayed on node *C* will be dropped if it follows the rule of perimeter forwarding strategy without position prediction, which would cause high end-to-end delays, low packet delivery ratio and high counted hops. By the predictive perimeter forwarding strategy, the route will be re-calculated based on predicted positions of the vehicle for packet transmission from source node *A*. Alternatively, source node *A* would deliver the packet by the new route $A \rightarrow B \rightarrow D \rightarrow E \rightarrow Z$.

In real-world scenarios, the speed and acceleration of a vehicle would have major impacts on its future positions. These two parameters can be easily extracted from both the onboard unit and the simulation. Therefore, the speed and acceleration ought to be used to predict the future position of the vehicle during the fixed communication interval. In comparison to the static geographic routing protocols, the prediction based solution can be more suitable and more realistic in a real-world scenario. Herein, the beacon messages containing the velocity, acceleration and position can be used by the source node to select the best neighbour to forward packets.

Consequently, the developed TDMP routing protocol combines the two predictive forwarding strategies discussed above. Instead of the static mechanism, TDMP uses the dynamic information of the vehicle and exchanges beacon messages to predict the position of the vehicle in a short-time window. The dynamic topology is more reasonable for transmitting the packet.

When deciding the next-hop neighbour to forward the packet, particularly, in some general scenarios, the travelling direction of a vehicle should not be ignored. When we look into the example in Fig.6, node *B*, *C* and *D* in the one-hop radio range of source node *A* have the same distance towards destination node *Z*. If we follow the rule of the greedy forwarding strategy, it would be stuck in the dilemma to select the best neighbour. However, when taking the direction of the vehicle into account, it will find node *C* is the most suitable next-hop neighbour.

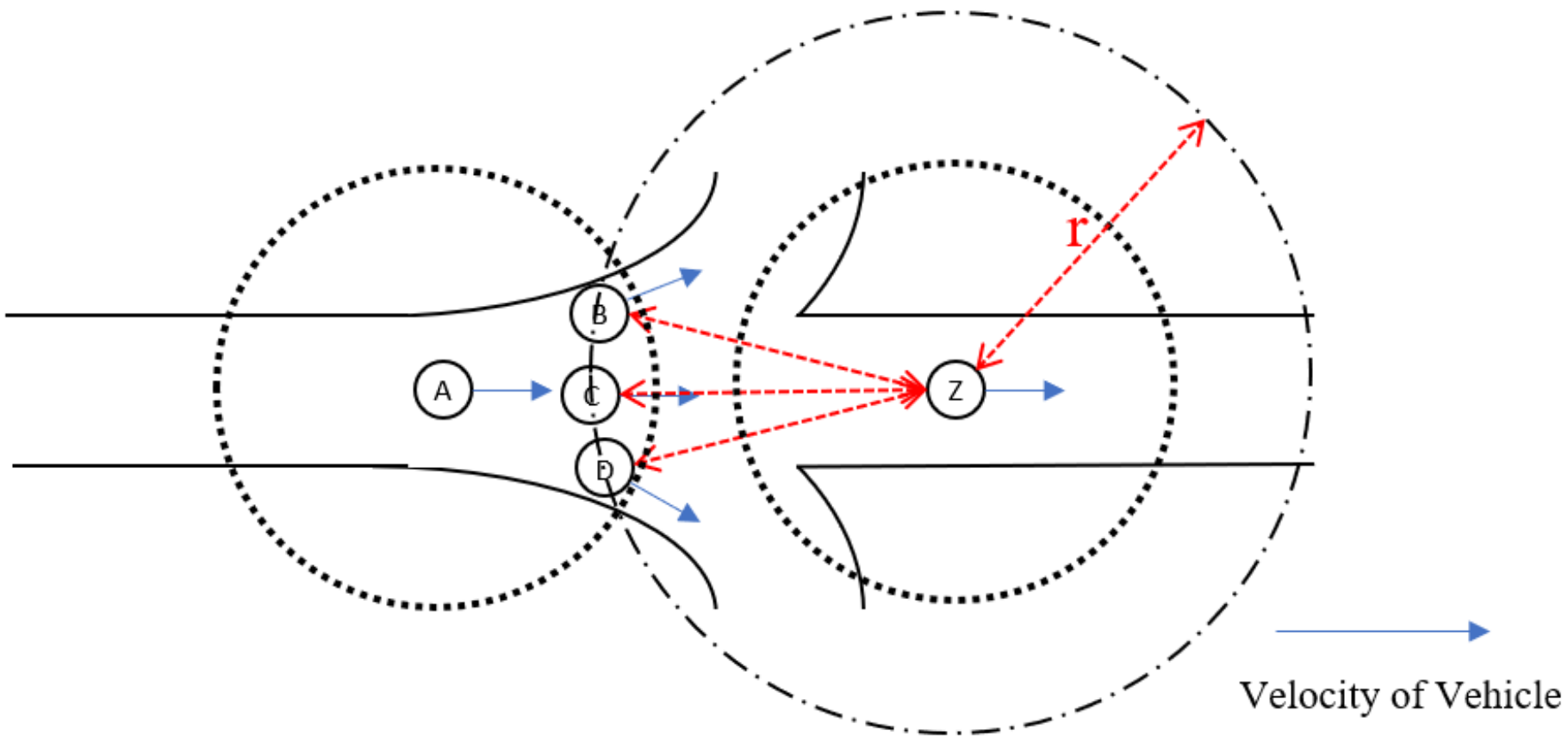

Fig.6. Direction Selection Scenario

Currently, most of the drivers tend to use the navigation and route guidance system before and during the journey to find the routes or to observe the traffic status, which can be described as the driver's target. For example, in terms of speed and acceleration, there is very little difference between node *B* and node *C*. Based on the predictive forwarding strategy, node *B* will be selected by source node *A* to relay the packet to destination node *Z*. However, the TDMP can measure the target of the vehicle, node *C* and destination node *Z* have a similar target in the scenario shown in Fig.7. Therefore, more weight will be given to node *C* to be selected as the relaying node to forward the packet. In particular, the target will take a significant impact on selecting the next-hop in some complex and irregular urban environments. In view of the high vehicle mobility in VANET, it should be noted that the position of each vehicle in Fig.6 and Fig.7 is predicted by source vehicle *A*.

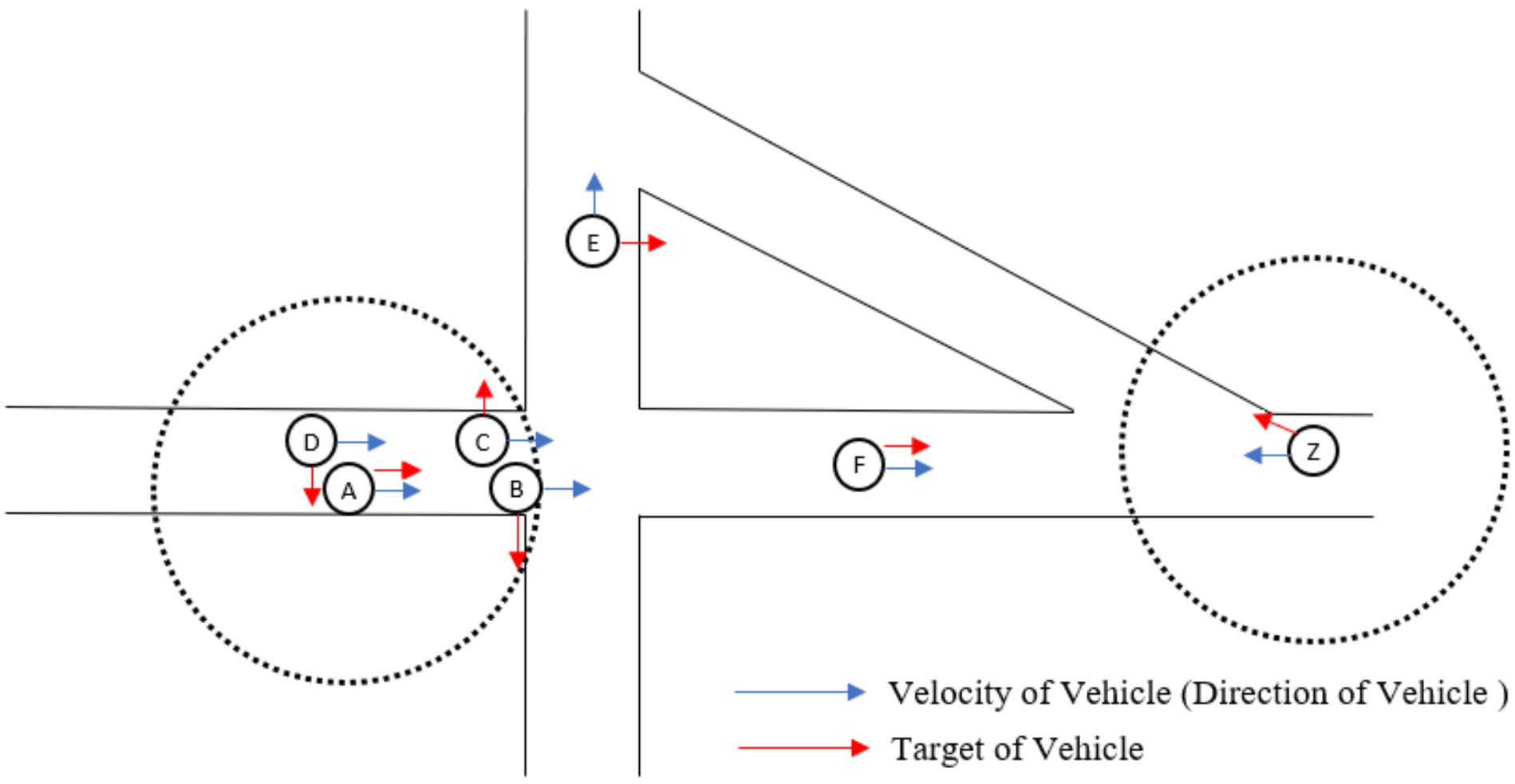

Fig.7.Target Driven Scenario

To formulate the factors inside the TDMP, some information should be calculated for the source node: the predicted Euclidean distance towards the destination of each neighbour, the predicted directional angle with the destination of each neighbour and the predicted target angle with the destination of each neighbour.

Before selecting the next-hop, the source vehicle would form the Potential Forwarders Group (PFG) which includes the neighbours that fulfil the following conditions and then evaluate the link status in a V2V environment.

- The average RSSI value obtained by calculating the RSSI from the neighbour must be greater than the predefined threshold as formulated in Equation (1). The RSSImax value is the maximum value among all the recorded RSSI values of the received beacons and data packets from all neighbours, where RSSI_threshold=0.6 x RSSImax.
- The predicted position of a neighbour is in the source vehicle transmission coverage area. The predicted position is calculated based on the mobility information (i.e. velocity, acceleration and recent position), which is obtained through beacons.
- The predicted position of the neighbour is closer to the destination than the source vehicle. If not, the recovery mechanism would be used to avoid trapping to a local maximum.

$$RSSI_Neighbour \geq RSSI_threshold \tag{1}$$

As a result, more priority is given to neighbours with high RSSI. The beforehand RSSI measurement can effectively mitigate the network overhead in a VANET environment.

Abovementioned factors would influence the next-hop selection by divergent weights. The framework of the developed TDMP routing protocol is presented in Fig.6 and there are a total of nine sequential steps as explained below:

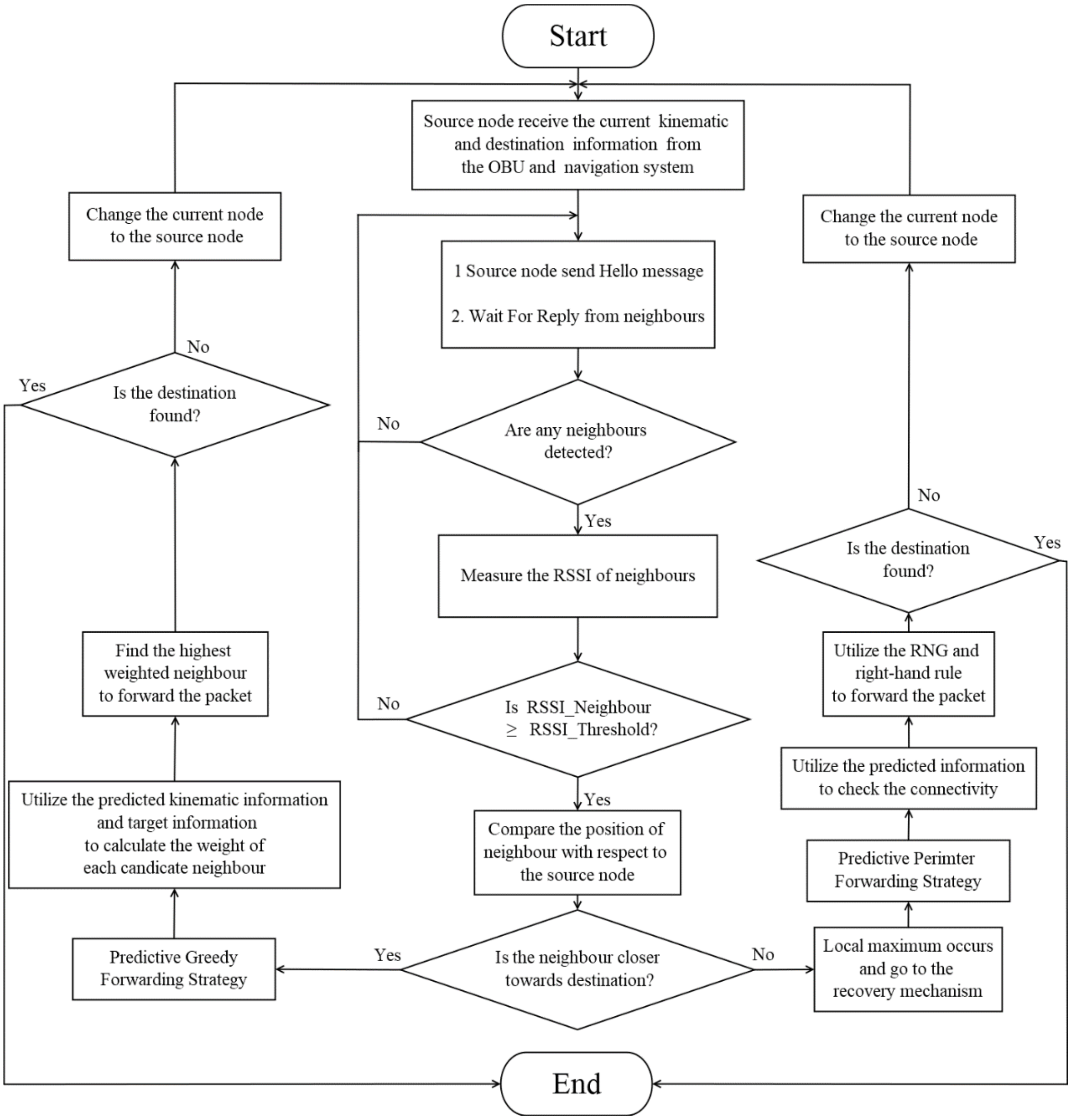

Fig.8. The framework of the TDMP routing protocol

Step 1: Receive the current kinematic information and the destination information from the OBU and the navigation systems.

Step 2: Send 'hello' message and detect any neighbours.

Step 3: Measure the RSSI value to evaluate the link status in V2V.

Step 4: Calculate the predictive Euclidean distance of each neighbour towards the destination node within the transmission range.

Step 5: Check whether the local maximum has arisen. If yes, go to the recovery mechanism (i.e. the predictive perimeter forwarding strategy). If not, go to the following steps.

Step 6: Calculate the predictive angle between the direction of each neighbour and the destination node within the transmission range.

Step 7: Calculate the predictive angle between the target of each neighbour and the destination node within the transmission range.

Step 8: Combine diverse factors and calculate the weight of each candidate neighbour.

Step 9: Select the best neighbour to deliver the messages.

To calculate the weight of neighbour nodes, three factors should be assigned to the relative distance and angles, $p$, $q_1$ and $q_2$. $p$ is the weighting factor for the distance, then $q_1$ and $q_2$ are the weighting factors for the two angles. The sum of $p$, $q_1$ and $q_2$ is 1.0, which are represented in Equation (2):

$$p + q_1 + q_2 = 1.0 \tag{2}$$

Ls $(X_s, Y_s)$ and Ld $(X_d, Y_d)$ denote the location of the source node and the location of the destination node respectively. The distance between the source node and the destination node is shown in Equation (3).

$$distance_{(s,d)} = \sqrt{(X_s - X_d)^2 + (Y_s - Y_d)^2} \tag{3}$$

The angle θ between the source node and the destination node is shown in Equation (4).

$$\theta = cos^{-1}\left(\frac{\vec{v} \times \overrightarrow{LsLd}}{\|\vec{v}\| \times \|\overrightarrow{LsLd}\|}\right) \tag{4}$$

where $\vec{v}$ denotes the vector of velocity for the source vehicle.

By the beacon interval *t*, the predicted position $(x', y')$ of each node can be calculated below. Equation (5) shows the predictive position of one node. Here, *v* is the velocity and *a* is the acceleration of the vehicle.

$$\begin{cases} x' = x + \left(vt + \frac{a}{2} \times t^2\right) \times cos\,\theta \\ y' = y + \left(vt + \frac{a}{2} \times t^2\right) \times sin\,\theta \end{cases} \tag{5}$$

Then, equation (1) (5) can be used to calculate the predictive $distance^*$ between two new nodes, which is shown in Equation (6).

$$distance^* = \sqrt{\begin{array}{c}\left(x_1 + \left(v_1 t + \frac{a_1}{2}t^2\right) cos\,\theta_1 - x_2 - \left(v_2 t + \frac{a_2}{2}t^2\right) cos\,\theta_2\right)^2 \\ + \\ \left(y_1 + \left(v_1 t + \frac{a_1}{2}t^2\right) sin\,\theta_1 - y_2 - \left(v_2 t + \frac{a_2}{2}t^2\right) sin\,\theta_2\right)^2\end{array}} \tag{6}$$

Where in Equation (6), $a$ denotes the acceleration of the vehicle and $\theta$ denotes the direction of the vehicle. $v_1$and $v_2$ are original speeds of two vehicles. $(x_1, y_1)$ and $(x_2, y_2)$ represent original positions of two vehicles respectively.

Next, Li $(X_i, Y_i)$, La $(X_a, Y_a)$ and Lb $(X_b, Y_b)$ denote the predicted location of neighbour, and the locations of neighbour's target and destination node's target respectively. The angle of the two targets can be calculated in Equation (7).

$$\theta i = cos^{-1}\left(\frac{\overrightarrow{L\iota La}\overrightarrow{LdLb}}{\|\overrightarrow{L\iota La}\| \times \|\overrightarrow{LdLb}\|}\right) \quad (7)$$

Finally, the weight for each neighbour *i* can be calculated by Equation (8). The neighbour with the higher weight can be the candidate for the next-hop.

$$weight_i =$$

$$p\left(\frac{distance^*_{s,d} - distance^*_{i,d}}{distance^*_{s,d}}\right) + q_1\, cos\left(\overrightarrow{v_\iota}, \overrightarrow{L_\iota L_d}\right) + q_2 cos\theta i \quad (8)$$

where $\overrightarrow{v_\iota}$ denotes the vector of velocity for the neighbour *i*.

Form (8), if a neighbour is closer to the destination with the similar direction as the destination and the intentions of them are similar, this neighbour can be given higher weight to be selected for relaying the packet.

3.3 Performance Metrics based on Quality of Services

Several popular geographic routing protocols in VANET are compared with TDMP, such as GPSR, GyTAR and PGRP. The performance of geographic routing protocols can be analysed by diverse parameters, such as Hop Count, End-to-end Delays, Packet Delivery Rate, Overhead and Latency. Three common metrics are chosen to evaluate the performance of the TDMP routing protocol.

End-to-End Delays (E2E Delays): The average time taken for successful packet transmission from the source vehicle to the destination vehicle.

Packet Delivery Ratio (PDR): The ratio of the total number of packets received ($N_{received}$) by the destination node successfully to the total number of packets generated ($N_{generated}$) by the source node originally. PDR can both represent the efficiency of packet transmission and measure the loss rate of packets during the transmission. The PDR is defined as follows:

$$PDR = N_{received} / N_{generated} \quad (9)$$

Average Hops Count (AHC): The average number of hops taken to transmit packets from the source vehicle to the destination vehicle. The path length in VANET is the number of hops by which the packet traverses from the source to the destination. AHC represents the quality of a path used in packet transmission.

## 4. Performance evaluation and analysis

This section presents the evaluation of the developed TDMP routing protocol. First, the individual components of TDMP are briefly analysed and discussed. Then, the overall performance is evaluated using a unified simulation platform — Veins which is developed to examine VANET-related protocols. Also, Veins supports various radio propagation models from simple free space to more complicated models designed for V2V communication [53]. Apart from the TDMP routing protocol, we considered three other VANET routing protocols, which are slightly modified to have a fair comparison with TDMP, i.e., having the same communication configuration and the same traffic volume. Herein, we present the simulation configuration, evaluation methodology, evaluation metrics for comparing the different protocols, the detailed simulation analysis and results.

## 4.1 Simulation configuration

A comprehensive review and analysis of existing studies indicate that Veins is an adaptable platform to test the TDMP routing protocol presented earlier [54]. Veins is an open source framework for running vehicular network simulations, which is based on two well-established simulators: OMNET++ and SUMO [55]. Road networks exported from the OpenStreetMap (OSM) can be modified to meet the requirements of the real environment by Java-OSM. SUMO can convert the map from the OpenStreetMap to generate the traffic flow [56]. We use Original/Destination metrics to populate vehicular traffic representing three scenarios: (i) a hypothetical grid-based network with high density traffic (Fig.9), (ii) a hypothetical grid-based network with low density traffic (Fig.10) and (iii) a real-world network (Fig.11). Hypothetical networks were created through NetEdit and the real network was obtained from the OSM representing Loughborough Town, Leicestershire, UK and this will be employed to validate the performance of TDMP. For each of the scenarios, the origin and destination nodes for each of the simulated vehicles are available via O/D matrix (assumed for Scenario 1 and 2). Travel demand O/D matrix for Loughborough was obtained from Leicestershire Country Council and employed to calibrate the simulation model. Traffic parameters used in SUMO for Scenario 1 and 2 are presented in Tab.1. In Scenario 3, a digital map from the OpenStreetMap is modified to simulate the realistic traffic flow of Loughborough (low traffic density), as shown in Fig.11. The parameters in Scenario 3 are similar to those of Scenario 1and 2, but the area size is expanded to 4km × 6km.

Objective Modular NETwork testbed in C++ (OMNET++) is connected with SUMO by the Veins platform. OMNET++ is an extensible, modular, discrete, component-based C++ simulation library and framework, primarily for building network simulators, which can handle vehicular communication [57].

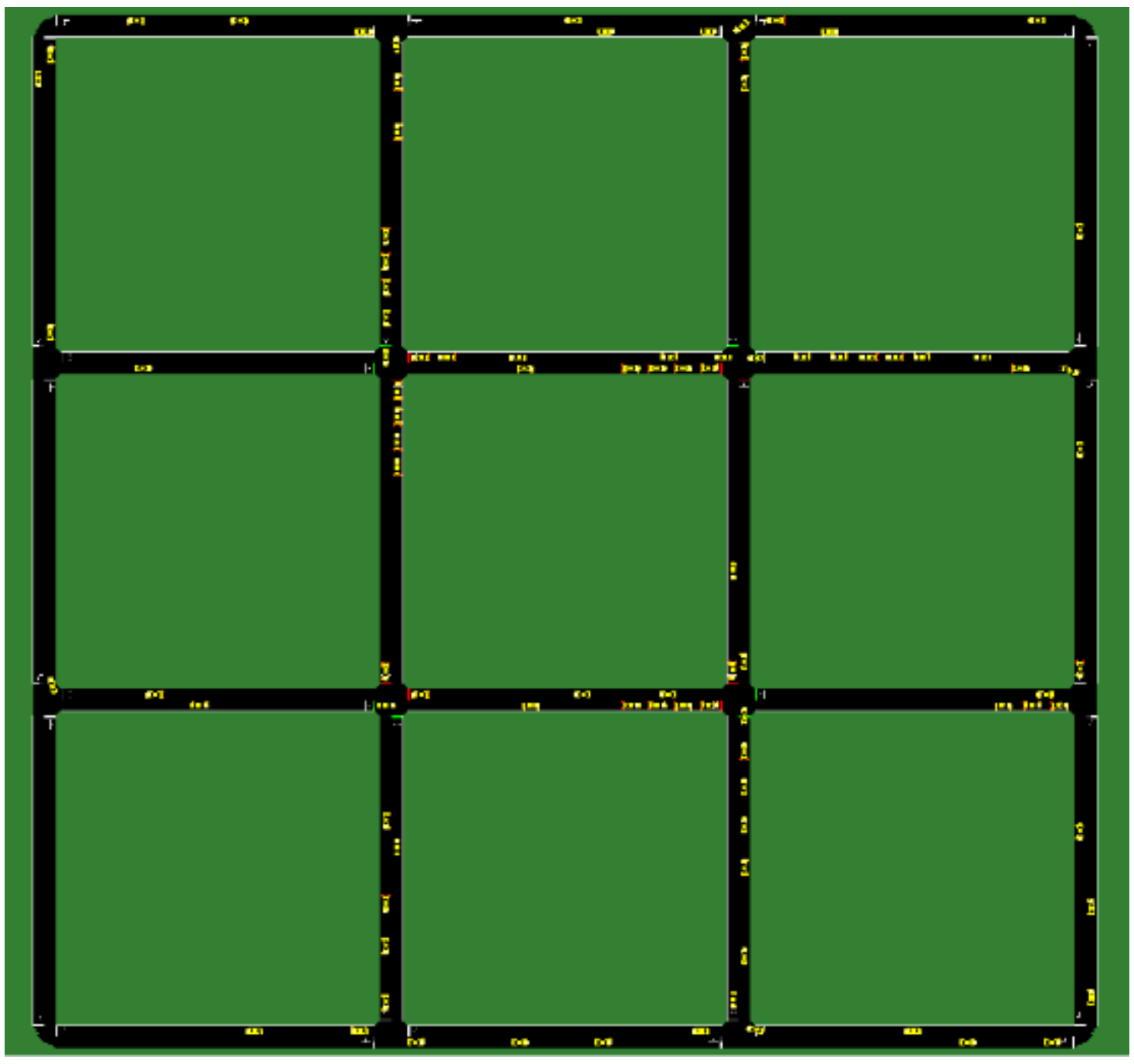

Fig9. The map of Scenario 1 (High-density traffic flow)

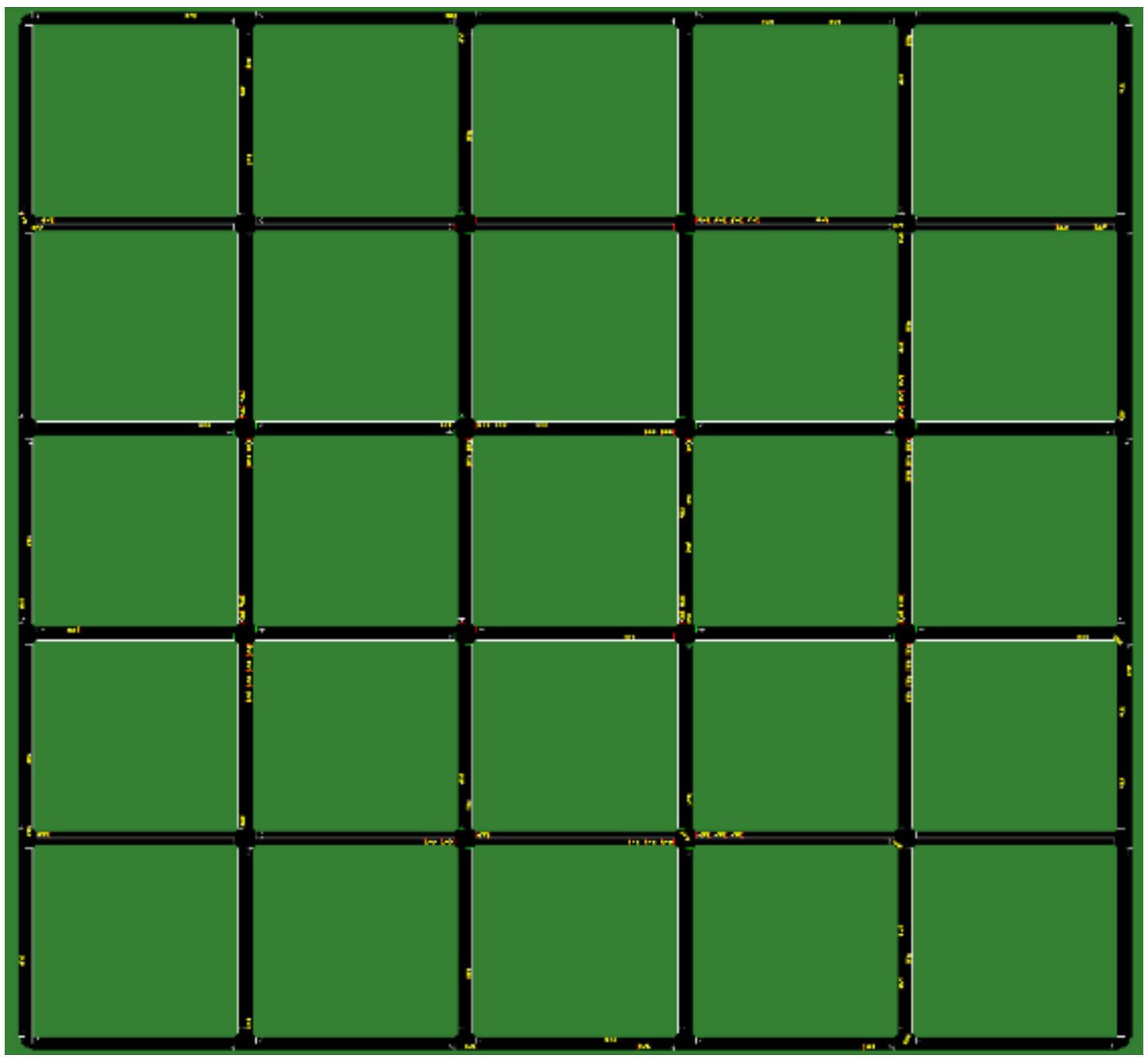

Fig.10.The map of Scenario 2 (Low-density traffic flow)

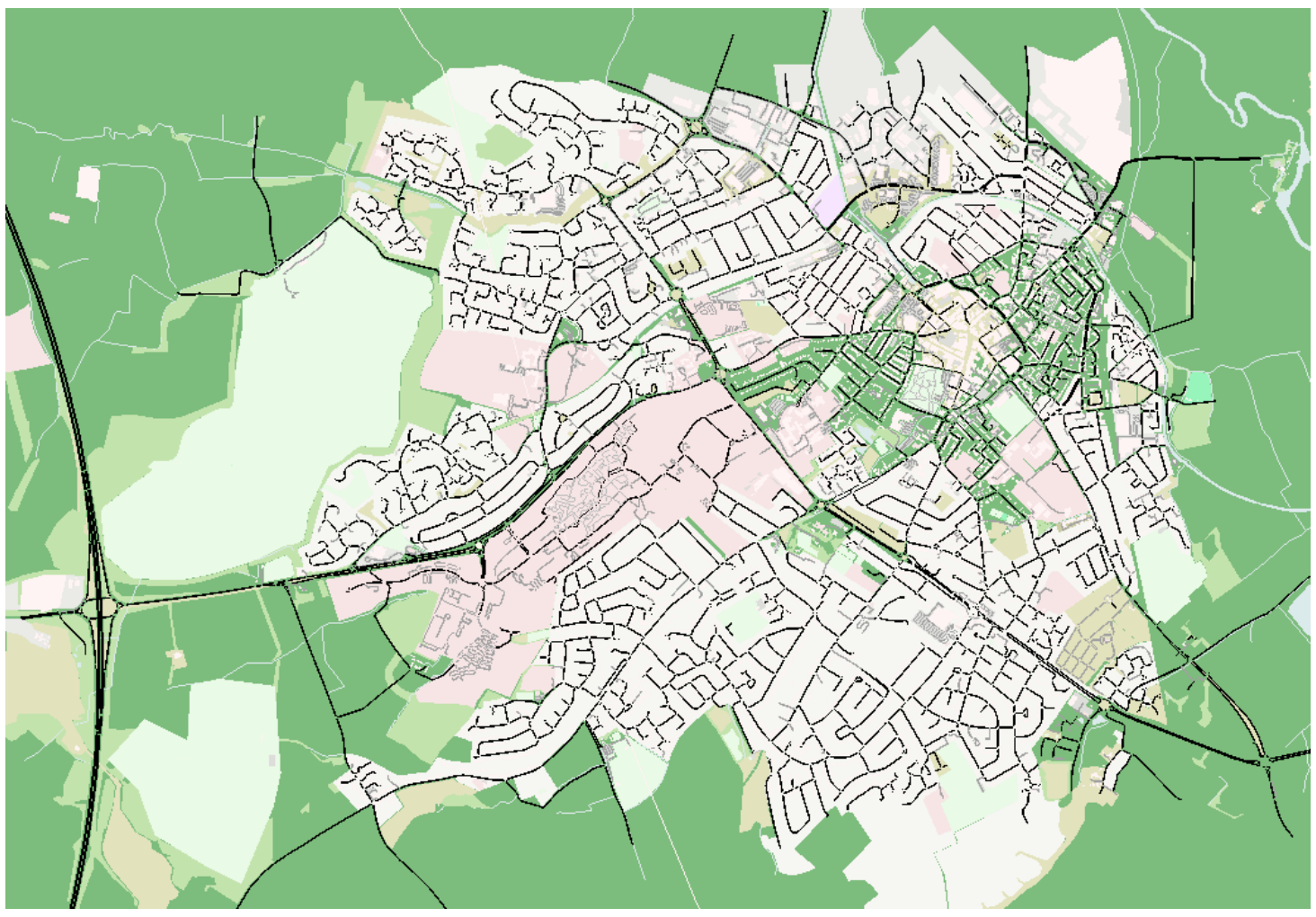

Fig.11.The map of Scenario 3 (Real-world map of Loughborough, UK)

Tab.1.Sumo parameters for grid-based scenarios

| SUMO parameters for grid-based scenarios | |
|---|---|
| Maximum vehicle speed | 13.3- 31.1 m/s (48.3 km/h -112.7 km/h) (30mph- 70mph) |
| Vehicle deceleration/acceleration | (-2.6- 4.5) m/$s^2$ |
| Vehicle length/ Width | 5m/ 1.8m |
| Mobility model | Car following model/ Krauss |
| Simulation network | 450 m × 450 m & 750 m × 750 m |
| Simulation time | 400s |
| Number of vehicles | 200, 400, 600, 800, 1000 |

The stochastic process of generating traffic flow depends on: (i) the O/D demand matrix and (ii) car-following model. The starting point, terminal point and traffic demand are saved in the O/D matrix. Within a given time period, the departure time of each vehicle follows a random distribution. In addition, the path of each vehicle is computed by SUMO by the well-known Dijkstra's shortest path algorithm. Based on the Dijkstra's algorithm, some numerical weights are associated with the corresponding road segments for route selections by the vehicles. In SUMO, the simulation updates the vehicles' positions in temporal steps of a user-specified duration $\Delta_t$ and move them by a positional increment $\Delta_x(t)$ under the rule of the first-order Euler scheme [56]. Meanwhile, the car following model describes the one-by-one following process of vehicles in a traffic stream and determines the safe speed of a vehicle in relation to the leading vehicle [58][59]. Herein, the safe speed is defined in the Krauss model as follows:

$$v_s = v_l(t) + \frac{d(t) - v_l(t)t_r}{\frac{v_l(t)+v_f(t)}{2a} + t_r} \tag{10}$$

where $v_l(t)$ denotes the speed of the lead vehicle at time t, $v_f(t)$ denotes the speed of the following vehicle at time t, $d(t)$ denotes the headway to the lead vehicle at time t, $t_r$ represents the reaction time of the driver and $a$ represents the maximum acceleration/deceleration of each vehicle.

Afterwards, Veins platform enables the communication process between the SUMO and OMNET++ to test the developed protocol. Relative parameters should be set up for the developed TDMP routing protocol, as shown in Tab.2. Radio propagation process as shown in Tab.2 follows the Two-Rays Ground Reflection Model, which simulates the path losses between a transmitting antenna and a receiving antenna in Line of Sight (LoS).

Tab.2.Veins parameters for three scenarios

| Veins parameters for the routing protocol- TDMP | |
|---|---|
| Beacon frequency | 1 Hz |
| Propagation model | Two-Rays Ground Reflection model |
| Transmission range | 300m |
| Channel capacity | 18 Mbit/s |
| Channel frequency | 5.89Ghz |
| Transmission power | 15mW |
| MAC layer protocol | IEEE 802.11p DSRC |
| Packet size | 1 KB |
| Sensitivity | -89 dBm |
| Weighting factor ($p$, $q_1$, $q_2$) | The detail in 4.2.(1) |

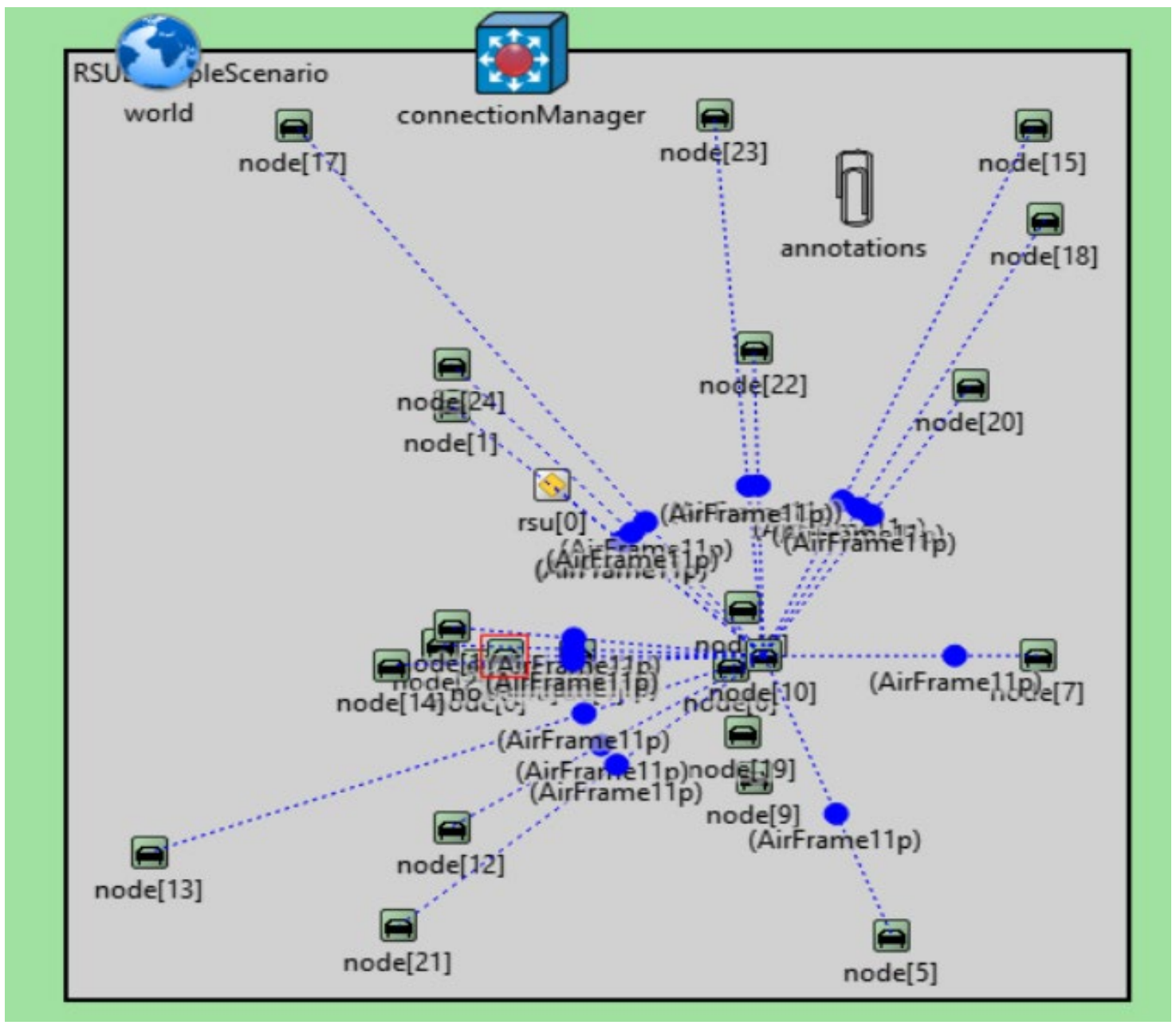

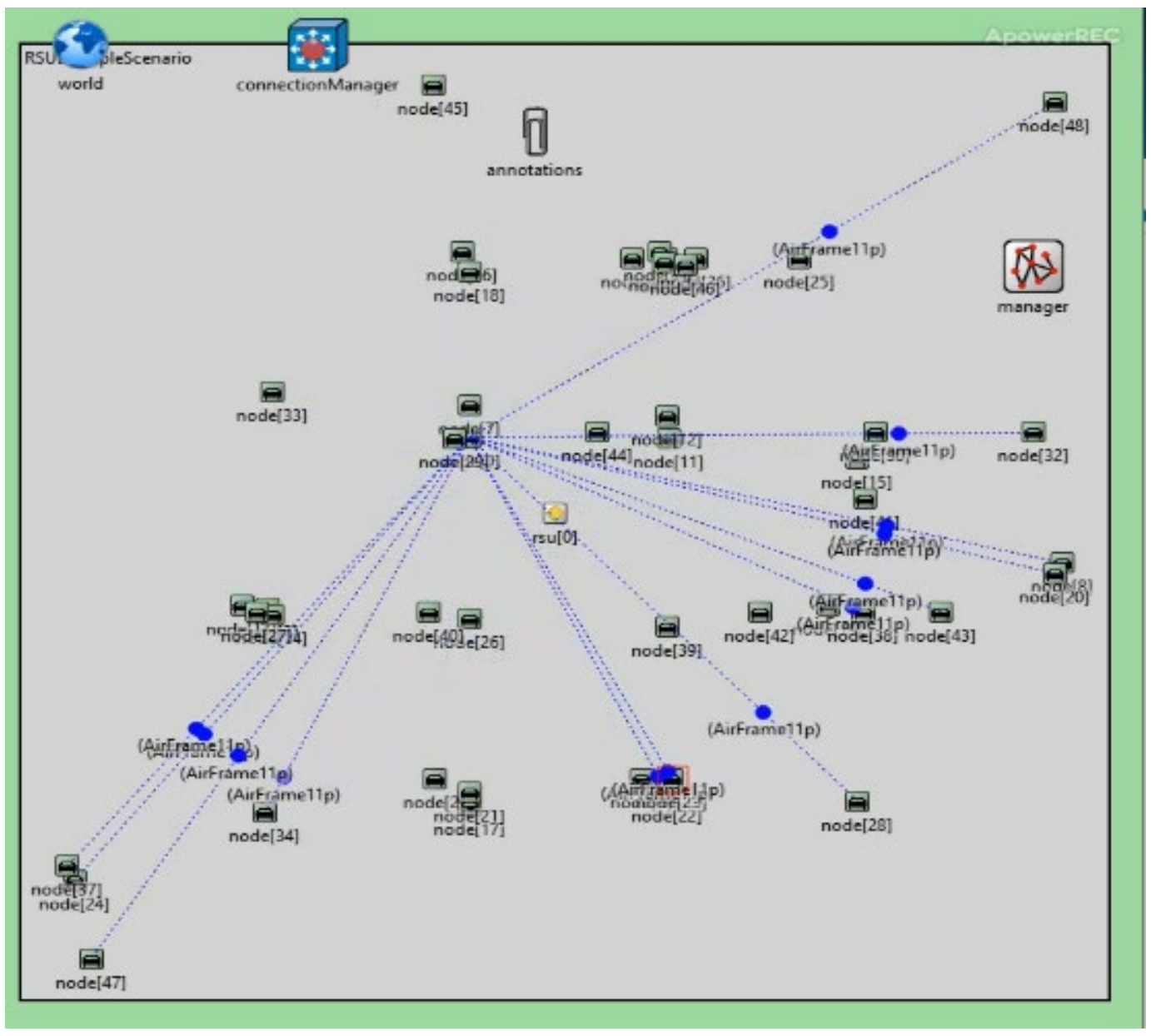

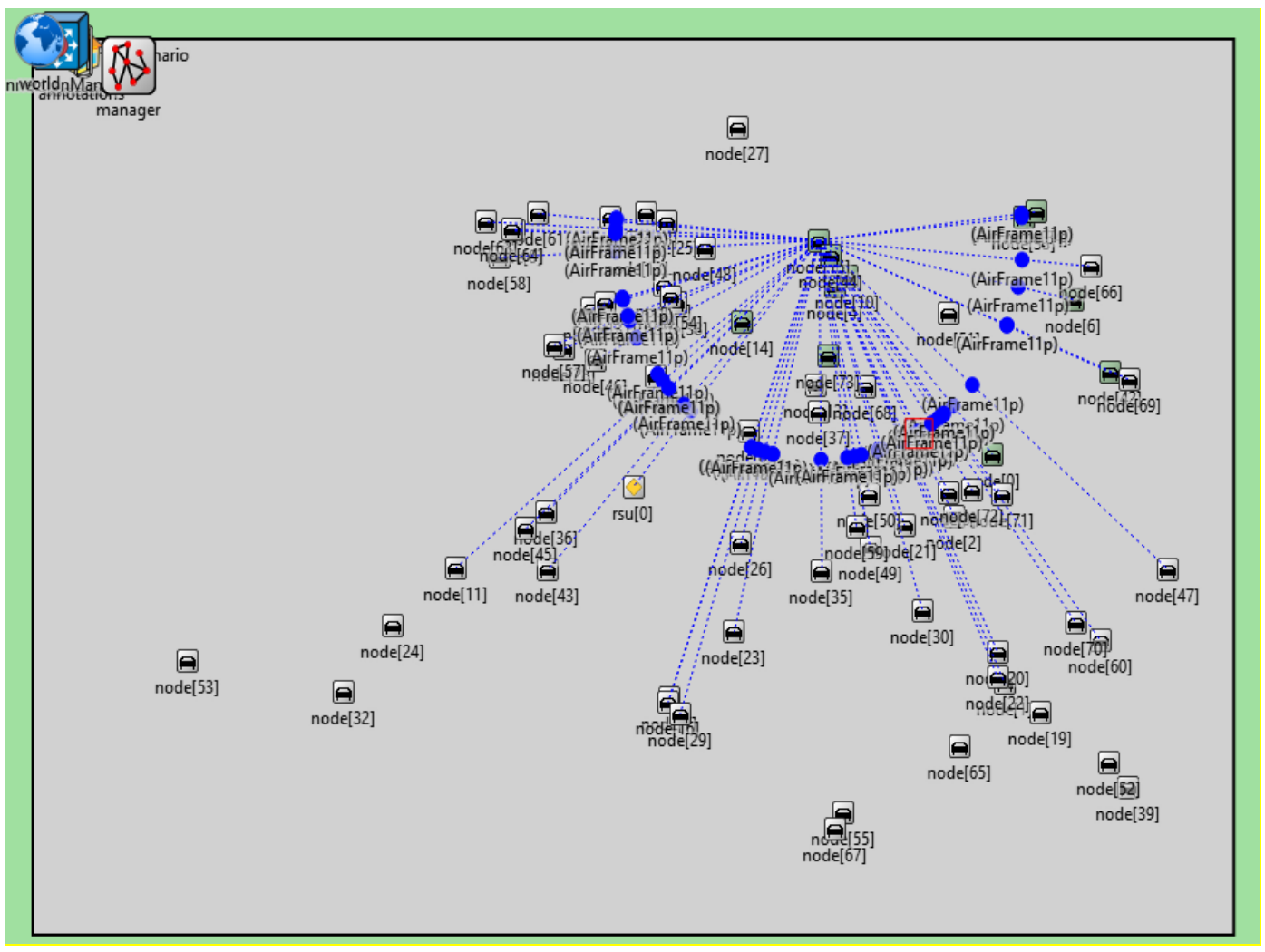

Fig.12. The packet transmissions of Scenario 1,2,3

4.2 Simulation results and analysis

1) Sensitivity analysis of the key parameters in different scenarios

This part presents the sensitivity analysis of the weighting factors in the fitness functions. Equation (8) chooses the candidate node for the next hop by combining the values of the distance to the destination, the angle to the destination and the intention of the driver based on the weights $p$, $q_1$ and $q_2$. Basically, the weighting factors are used to determine the percentage of contribution for each component in calculating the next hop. For instance, if the condition of equal importance is imposed then the weighting factors takes the following values: $p$=0.333, $q_1$=0.333 and $q_2$=0.333 meaning that each component of the fitness function contributes 33.3% in the calculation of the next-hop candidate. In general, there are no optimal ratios to be assigned to $p$, $q_1$ and $q_2$ which give the best routing performance in all scenarios.

Therefore, the sensitivity analysis for $p$, $q_1$ and $q_2$ ratio is conducted to determine the most suitable values for making routing decisions.

Although there could be uncountable numbers of combinations, four different weighting factors are evaluated to analyse the effect of $p$, q1 and $q_2$ values on the routing performance, where TDMP (0.333,0.333,0.333) corresponds to the configuration of $p$=0.333, $q_1$=0.333 and $q_2$=0.333, TDMP (0.4,0.3,0.3) corresponds to the configuration of $p$ =0.4, $q_1$=0.3 and $q_2$=0.3, TDMP (0.3,0.4,0.3) corresponds to the configuration of $p$ =0.3, $q_1$=0.4 and $q_2$=0.3 and TDMP(0.3,0.3,0.4) corresponds to the configuration of $p$ =0.3, $q_1$ =0.3 and $q_2$ =0.4. In TDMP (0.333.0.333,0.333), each component is equal. However, TDMP (0.4,0.3,0.3) favours the distance to the destination, TDMP (0.3,0.4,0.3) favours the angle to the destination and TDMP (0.3,0.3,0.4) favours the driver's intention while selecting the next-hop relay. The four considered configurations - TDMP (0.333,0.333,0.333), TDMP (0.4,0.3,0.3), TDMP (0.3,0.4,0.3) and TDMP (0.3,0.3,0.4) are evaluated in terms of packet delivery ratio (PDR) in both high and low vehicular density scenarios. Based on scenario 1 and scenario 2, the simulation results of PDR have been shown in Fig.13.

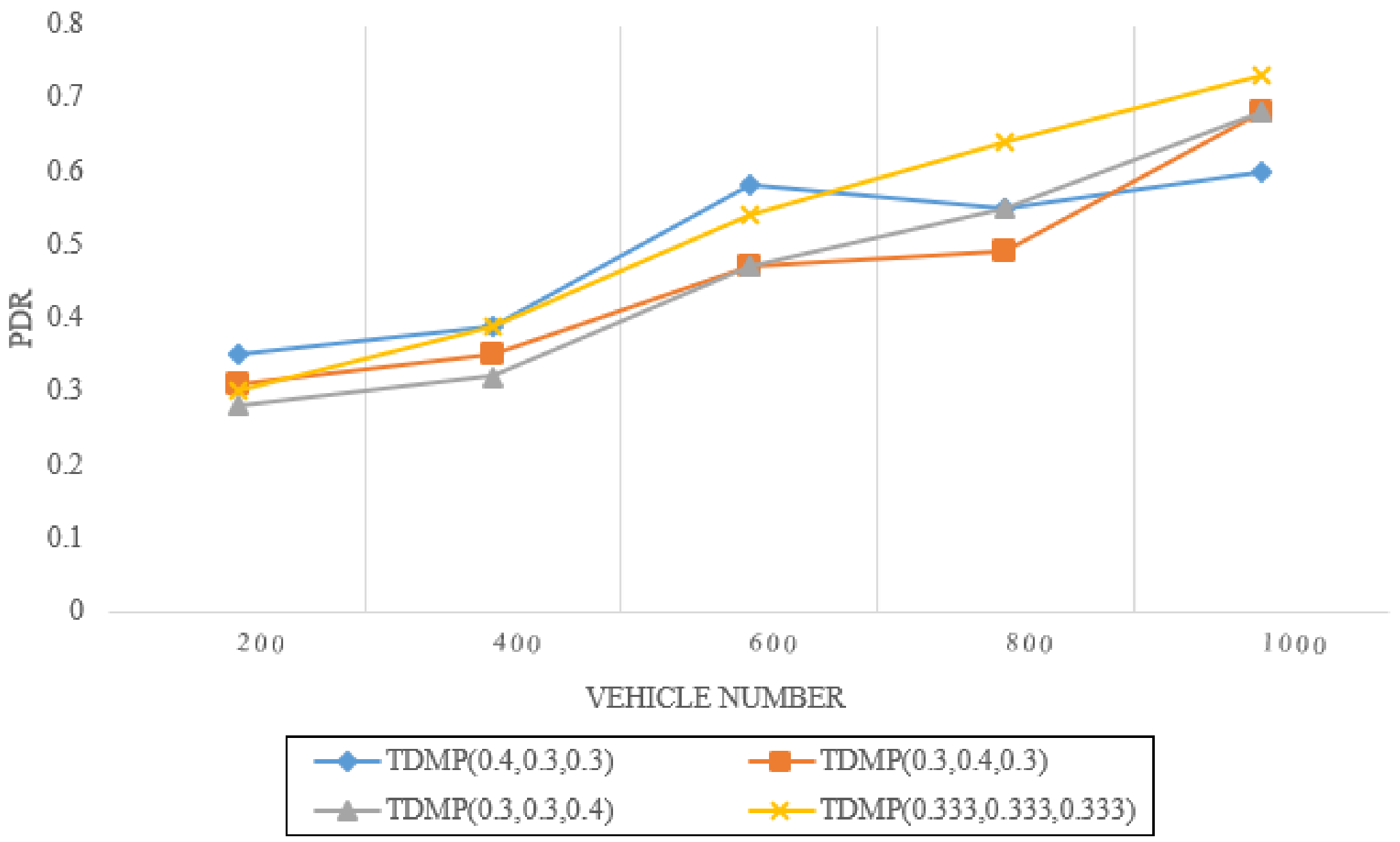

(a) PDR in scenario 1

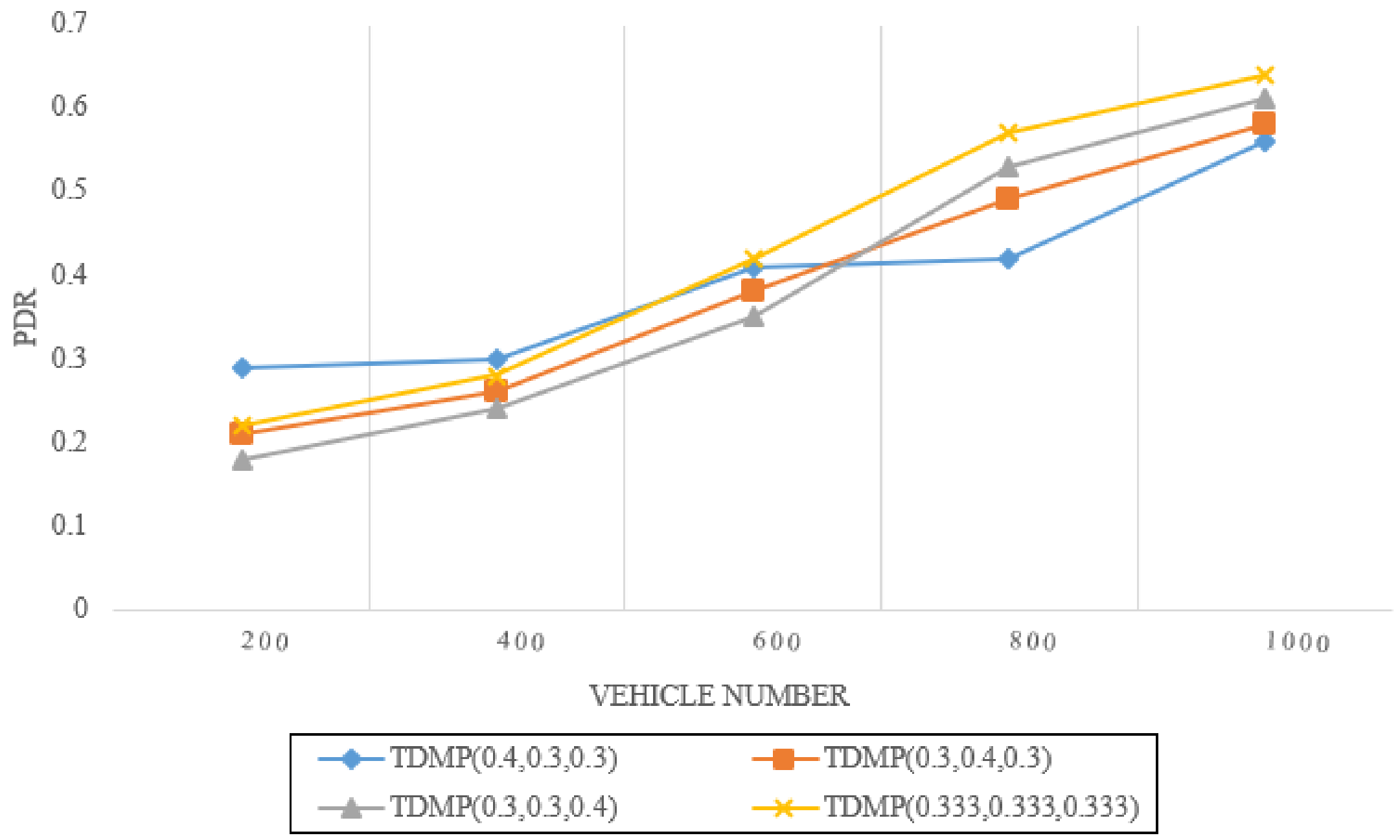

(b) PDR in scenario 2

Fig.13.PDR for TDMP in different parameter configurations

According to the results of Fig.13, it is obvious that each weighting factor has a positive and negative impact on the performance of the TDMP routing protocol in different vehicle number under different traffic density. Therefore, it is better to trade off between different weighting configurations to achieve a more balanced influence. Here, TDMP (0.333,0.333,0.333) is chosen to give equal importance for three factors in the fitness function. TDMP (0.333,0.333,0.333) is assigned uniform values for all subsequent simulations.

2) Analysis of TDMP in different scenarios

This section presents the performance evaluation of TDMP, which is compared against GPSR, GyTAR and PGRP respectively. The following three performance metrics have been introduced in Section 3.3, Packet Delivery Ratio (PDR), End-to-End Delay (E2ED) and Average Hop Count (AHC). Based on the different traffic patterns and vehicle density, the evaluation has been carried out in three different scenarios, which have been introduced in Section 4.1.

A. Scenario1

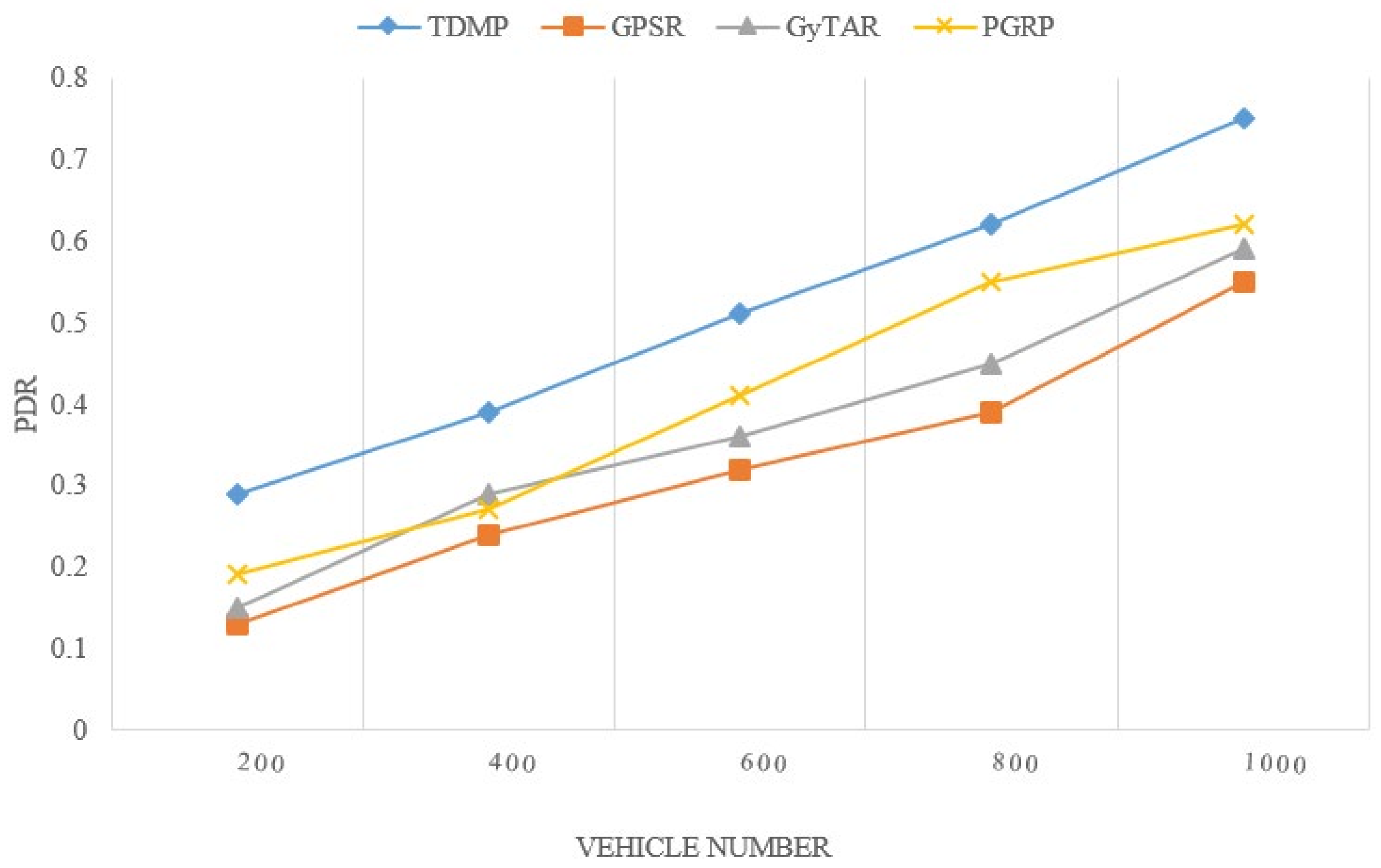

(a) PDR in Scenario1

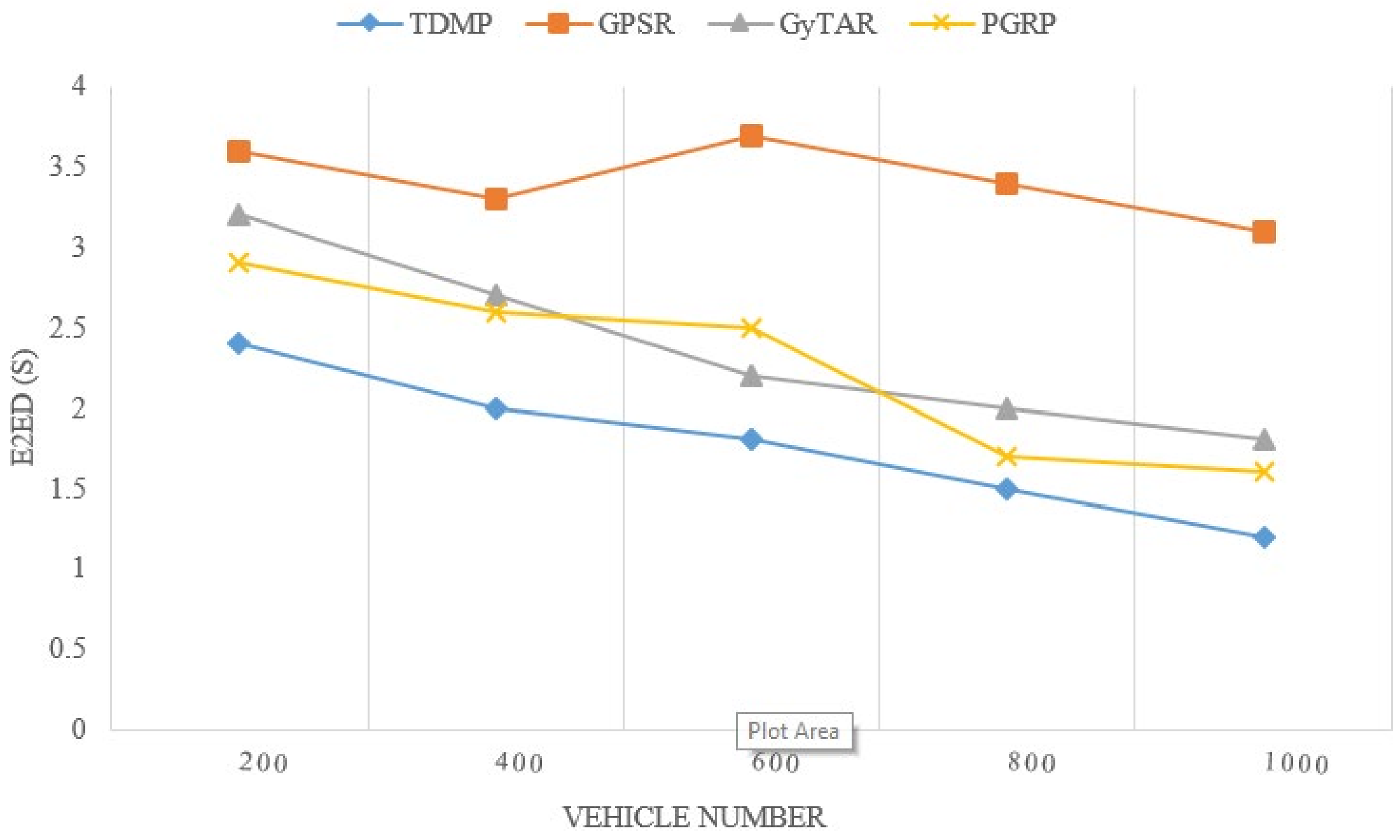

(b) E2ED in Scenario 1

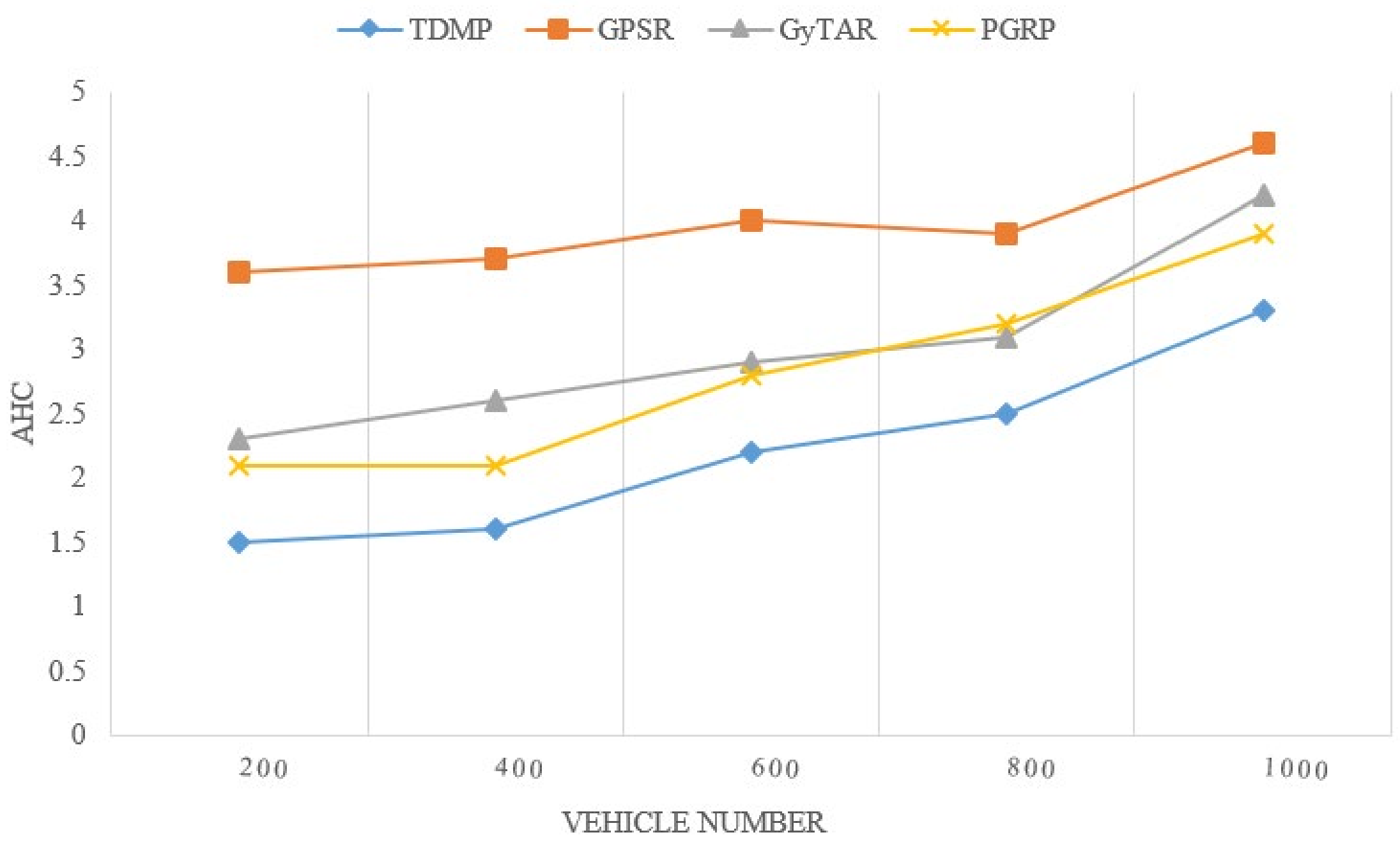

(c) AHC in Scenario 1

Fig.14. Results for Scenario 1

As shown in Fig.14, the developed TDMP achieved better results with respect to PDR, E2ED and AHC. Fig.14.(a) indicates that the TDMP's packet delivery ratio which is compared with GPSR, GyTAR and PGRP. As expected, with the increase of the vehicle density, the PDRs of TDMP, GPSR, GyTAR and PGRP all grow. However, TDMP outperforms GPSR, GyTAR and PGRP. As shown in Fig.14.(a), the PDR of TDMP is 57.06%, 39.13% and 25.49% higher than GPSR, GyTAR and PGRP's respectively. PGRP's result is close to TDMP's because both of these two routing protocols use the prediction method to choose the forwarding node. Fig.14.(b) shows their E2ED which are decreased with the growth of the number of the vehicle. And the E2ED of TDMP is 47.95%, 25.21%, 21.81% lower than GPSR, GyTAR and PGRP's respectively. Similarly, Fig.14.(c) shows that AHC also increases with the growth of vehicles. In particular, the AHC of TDMP outperforms GPSR, GyTAR and PGRP's by 47.95%, 25.21%, 19.09%.

B. Scenario2

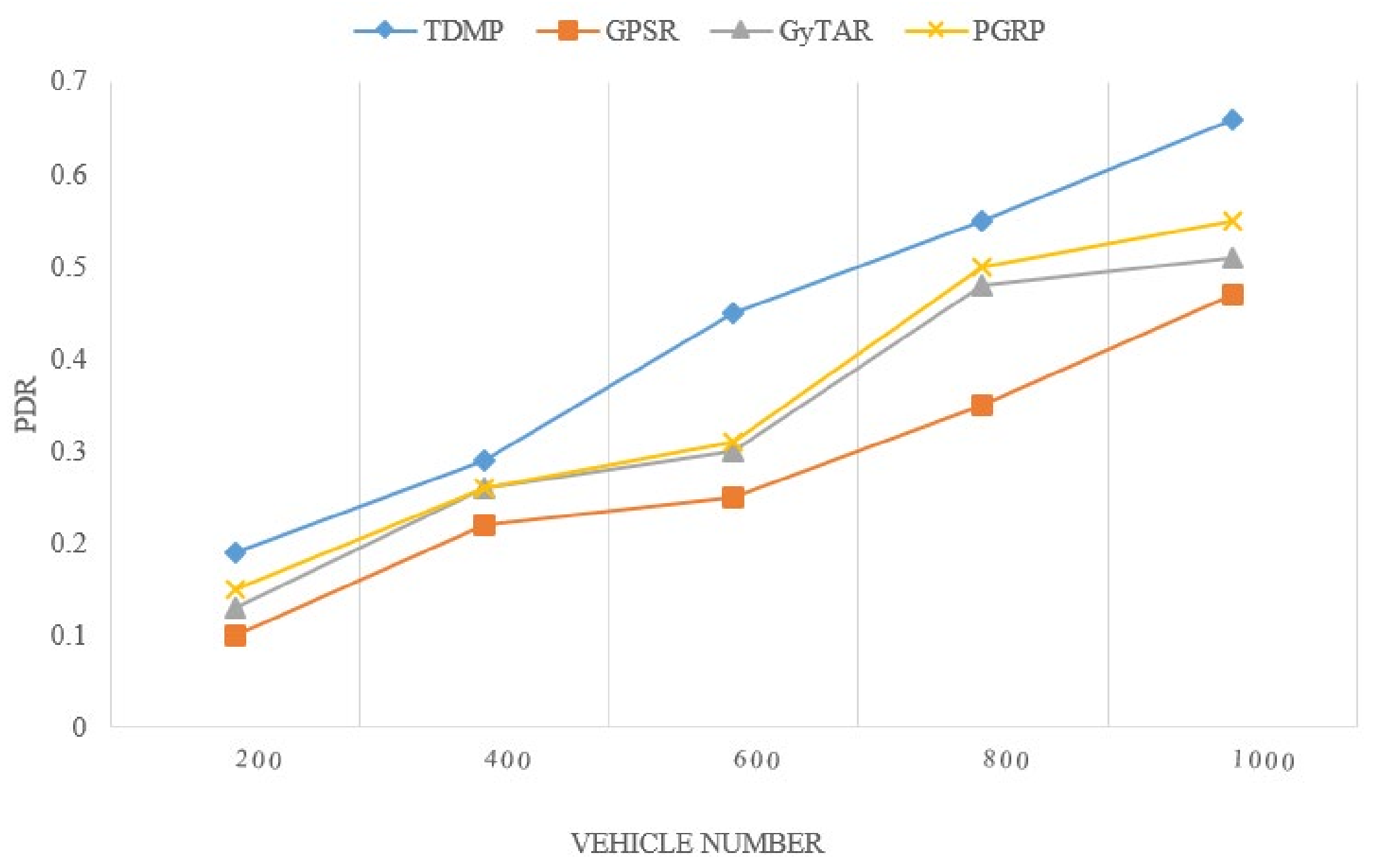

(a) PDR in Scenario 2

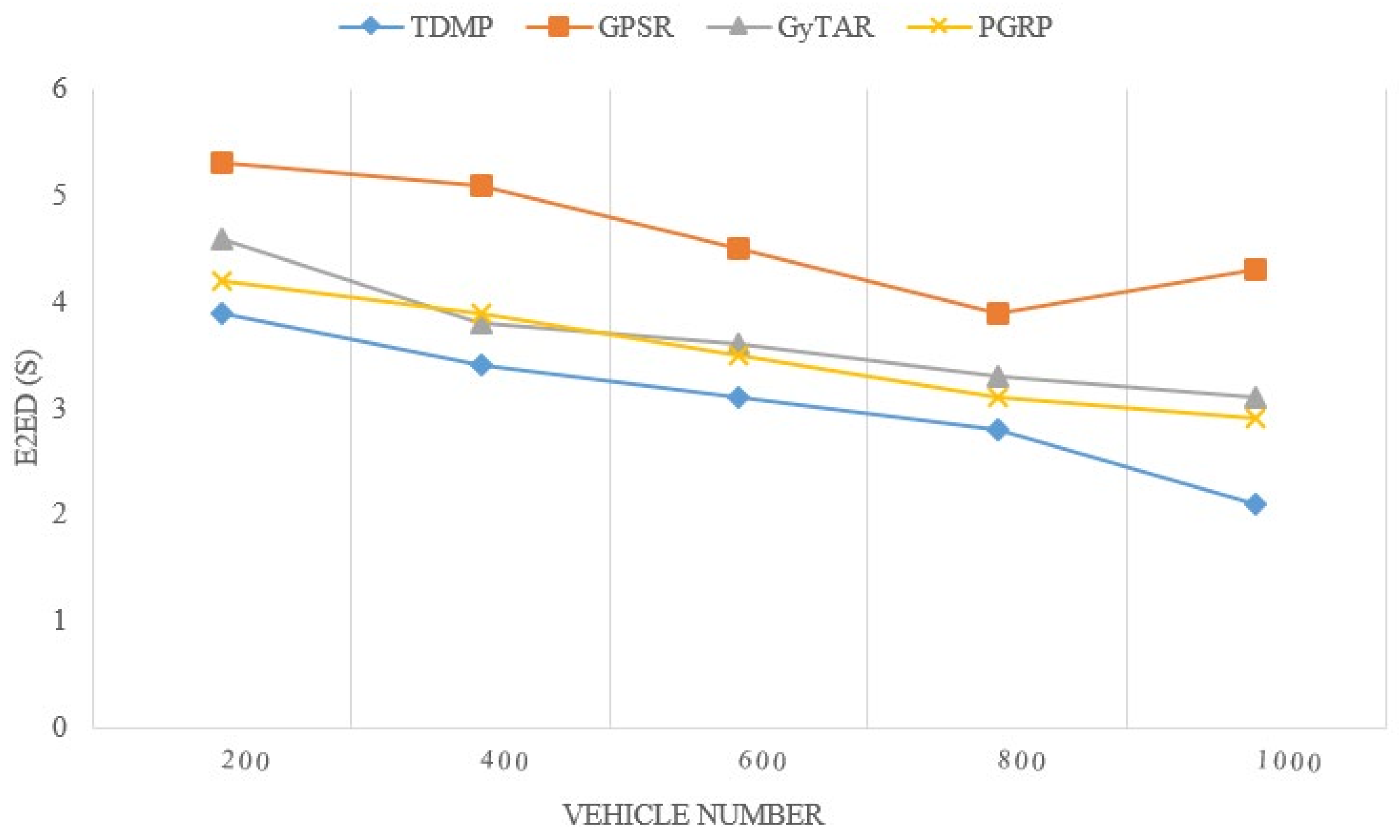

(b) E2ED in Scenario 2

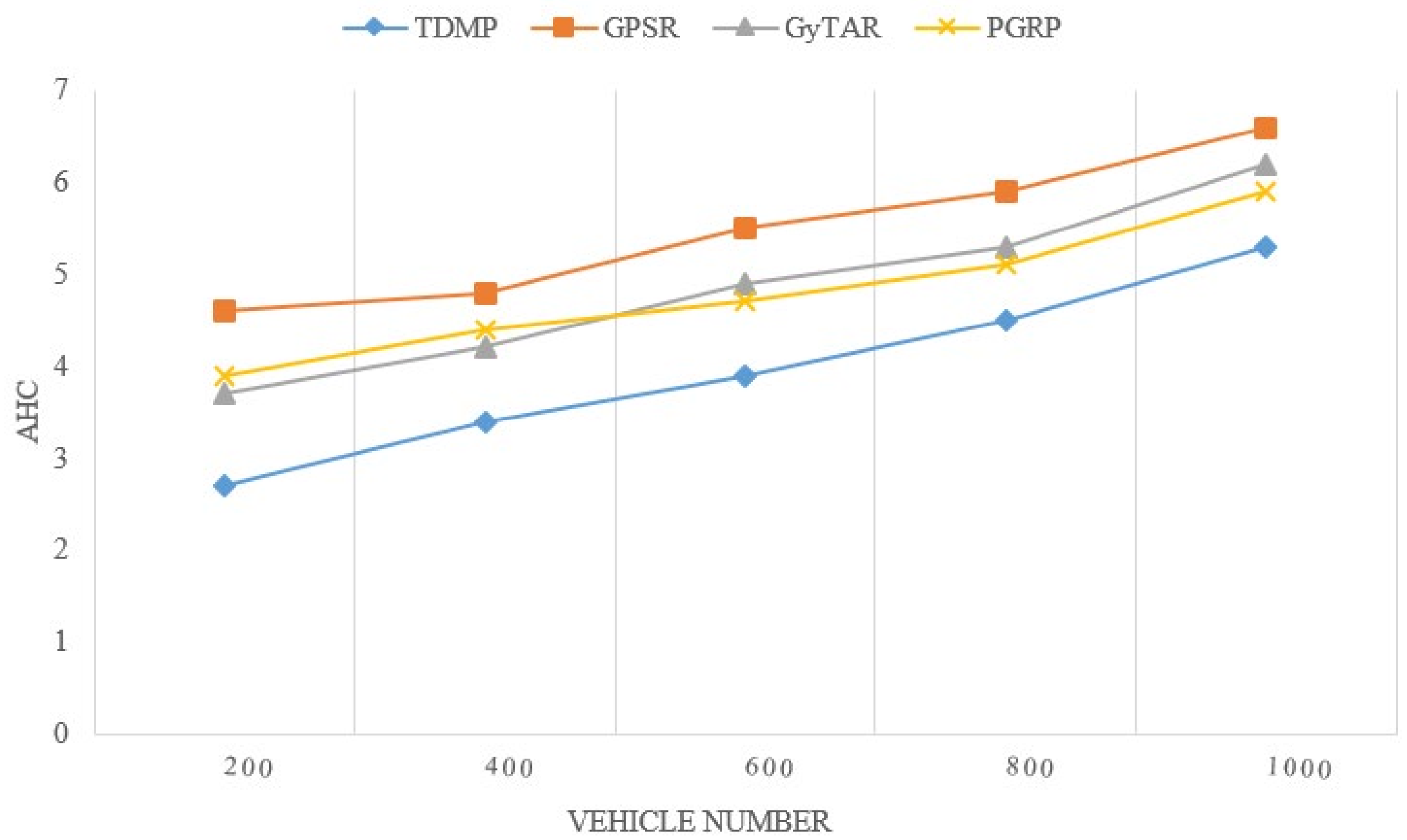

(c) AHC in Scenario 2

Fig.15. Results for Scenario 2

As shown in Fig.15, the developed TDMP performed better regarding PDR, E2ED and AHC. Fig.15.(a) indicates that the TDMP's packet delivery ratio which is compared with GPSR, GyTAR and PGRP. As expected, with the increase of the vehicle density, the PDRs of TDMP, GPSR, GyTAR and PGRP all grow. However, TDMP still outperforms GPSR, GyTAR and PGRP. As shown in Fig.15.(a), the PDR of TDMP is 53.96%, 27.38% and 20.90% higher than GPSR, GyTAR and PGRP's respectively. PGRP's result is still close to TDMP's because of the similar prediction mechanism for the packet routing. Fig.15.(b) shows their E2ED which generally decline with the growth of the number of the vehicle. And the E2ED of TDMP is 33.76%, 16.85%, 13.07% lower than GPSR, GyTAR and PGRP's respectively. Similarly, Fig.15.(c) shows that AHC also increases with the growth of the number of the vehicle. Particularly, the AHC of TDMP outperforms GPSR, GyTAR and PGRP's by 36.24%, 23.72%, 18.97%.

C. Scenario3

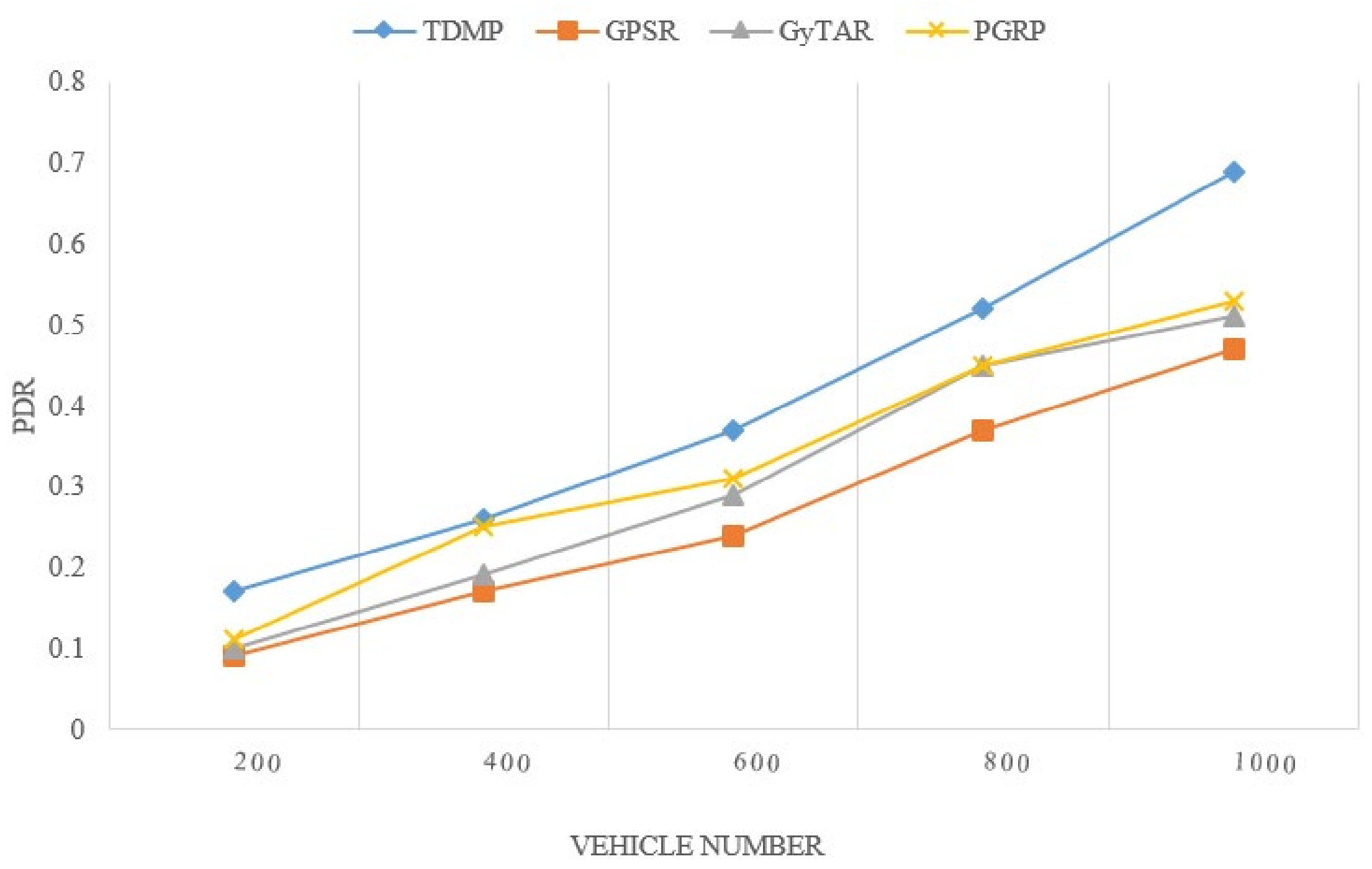

(a) PDR in Scenario 1

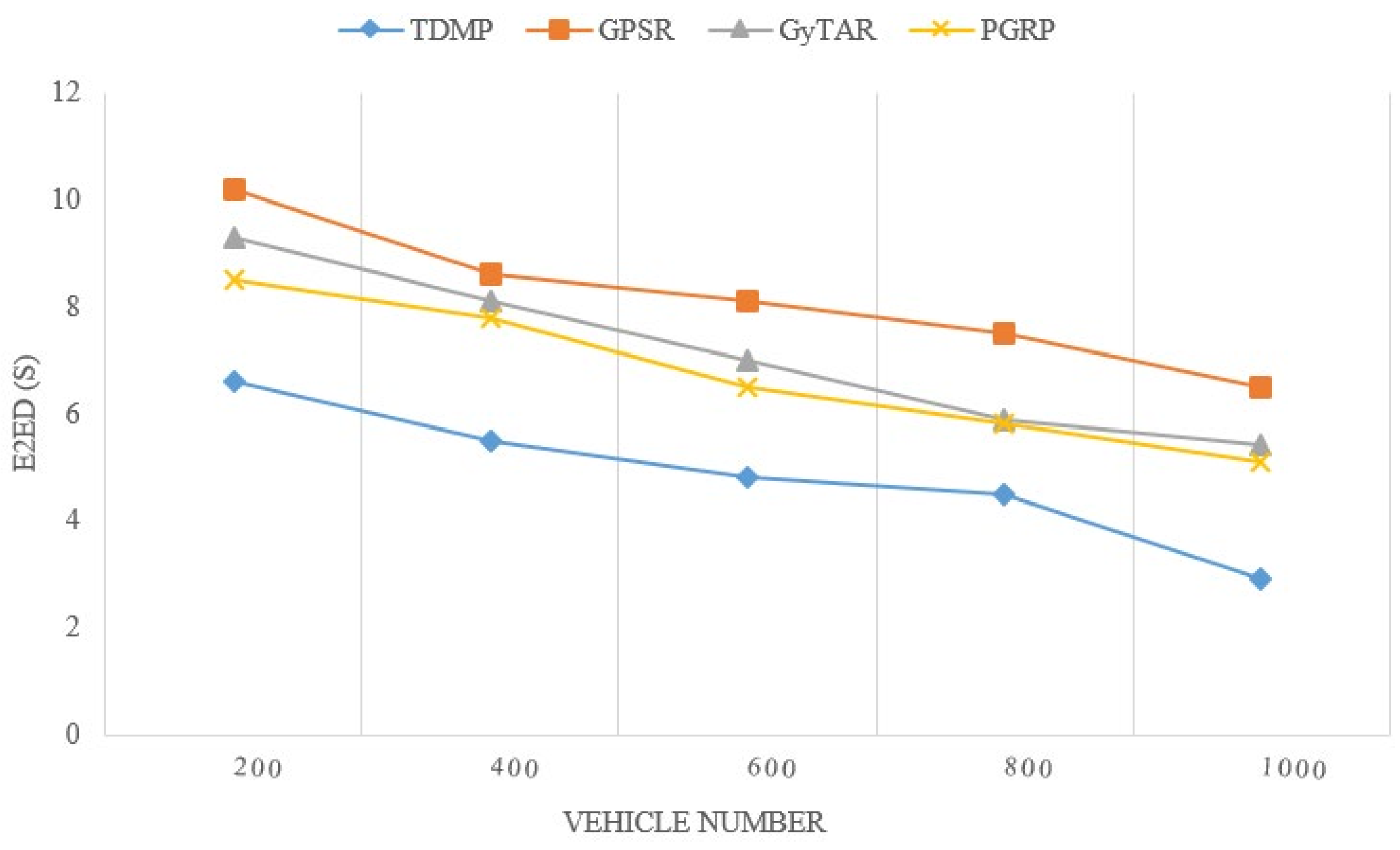

(b) E2ED in Scenario 2

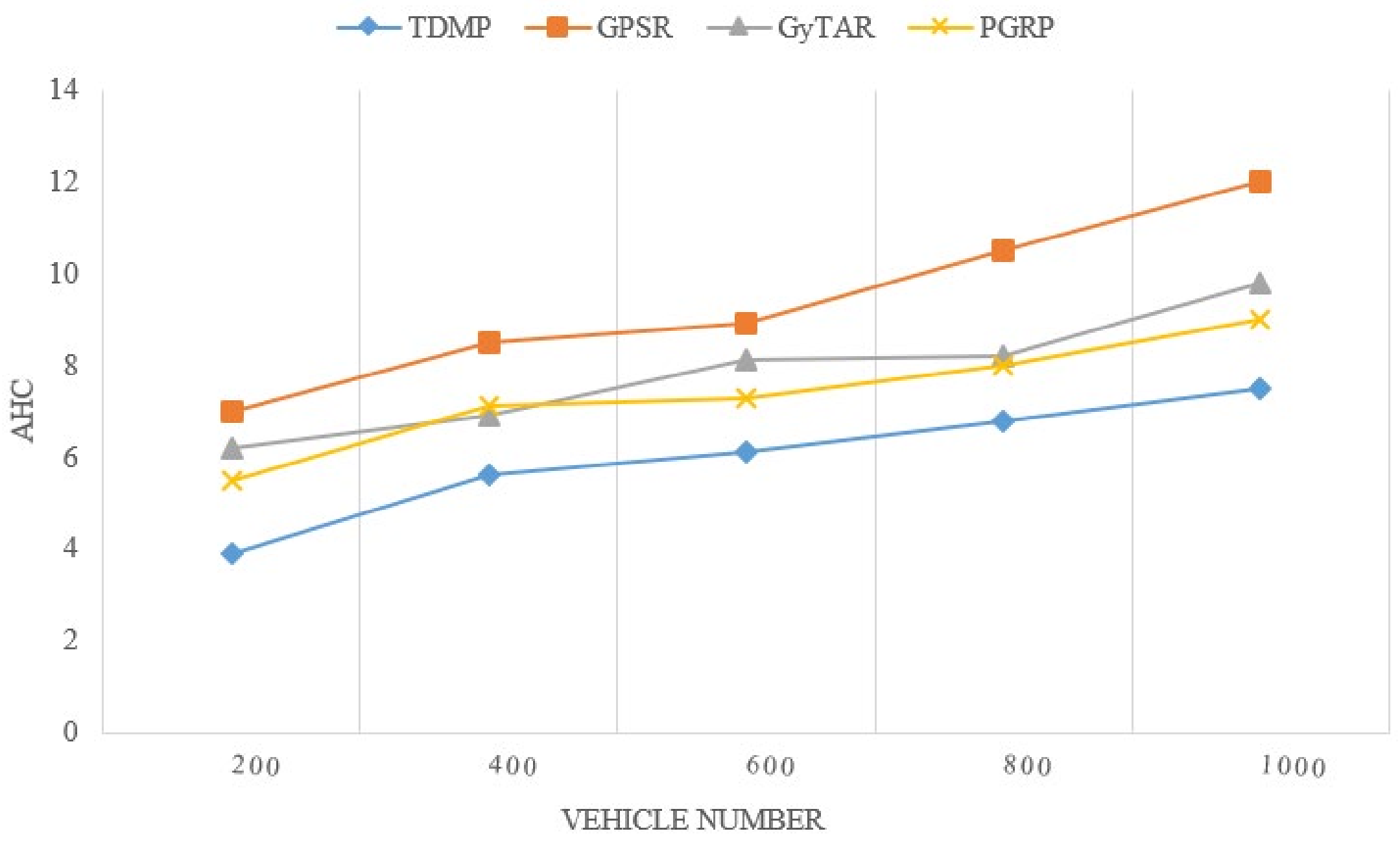

(c) AHC in Scenario 3

Fig.16. Results for Scenario 3

As shown in Fig.14, the developed TDMP achieved superior results in regards to PDR, E2ED and AHC. Fig.16.(a) indicates that the TDMP's packet delivery ratio which is compared with GPSR, GyTAR and PGRP. As expected, with the increase of the vehicle density, the PDRs of TDMP, GPSR, GyTAR and PGRP all grow. However, TDMP outperforms GPSR, GyTAR and PGRP. As shown in Fig.16.(a), the PDR of TDMP is 50.02%, 30.52% and 21.81% higher than GPSR, GyTAR and PGRP's respectively. PGRP's result is close to TDMP's because both of these two routing protocols use the prediction method to choose the forwarding node. Fig.16.(b) shows their E2ED which are decreased with the growth of the number of the vehicle. And the E2ED of TDMP is 37.41%, 28.29%, 24.04% lower than GPSR, GyTAR and PGRP's respectively. Fig.16.(c) shows that AHC also roughly increases with the growth of the number of the vehicle. In particular, the AHC of TDMP outperforms GPSR, GyTAR and PGRP's by 36.25%, 23.72%, 18.97%.

By comparing the performance of the developed TDMP with several commonly employed PBR protocols, it has been demonstrated that the TDMP offers significant improvements with respect to key performance indicators such as PDR, E2ED and AHC. The developed TDMP protocol not only integrates the advantages of GPSR, GyTAR and PGRP but also includes two important factors: (i) the measurement of the vehicle connectivity and (ii) the target of the vehicle. These improvements make TDMP superior to existing PBR protocols. By evaluating the performance in three typical scenarios, the following arguments can be drawn:

Firstly, the prediction mechanism has crucial effects on the performance of the VANET routing protocol, particularly in either a high dynamic topology or a high dynamic vehicle movement. This means that PBR protocols with the movement prediction achieve better results than ones without any prediction mechanism. Secondly, TDMP enables to achieve better performance when used in the scenario with higher traffic density because of the improved movement prediction and link-status measurement. Thirdly, TDMP can perform well in the complex urban environment because of the prediction and judgement of the targets of the vehicle as demonstrated in the real-world simulation of Loughborough town.

TDMP can be further improved by incorporating a new mechanism to trade off the factors of the current state and the predicted state. The prediction mechanism works well in VANET routing protocol. A balance must be maintained between the current estimation and the predicted estimation so as to achieve a precise decision for packet forwarding and routing.

## 5. Conclusion and future perspectives

In this paper, a new position based target-driven routing protocol termed as the Target-Driven and Mobility Prediction based routing protocol (TDMP) in VANET was developed. TDMP is capable of dealing with the inherent V2X challenges associated with heterogeneous traffic and complex urban environments. Relative to existing similar routing protocols, the developed TDMP achieved better performance by sensing the neighbouring environment, avoiding the local maximum, selecting the best next-hop node and checking the connectivity to the destination. Enhanced forwarding strategies support the TDMP with improving the packet delivery ratio by a margin of 21-57%, overcoming end-to-end delays by 13-47% and reducing the average hop by 17-48%. The simulation results showed that compared with GPSR, GyTAR, and PGRP, the developed routing protocol achieved competitive improvement in terms of packet delivery ratio, end-to-end delays and average hop count. By comparing the three scenarios, it was found that the results for small network scenarios performed better than the bigger network ones. The innovation aspect of the developed protocol relates to the utilisation of RSSI before the selection of a relay node, which guarantees that the vehicle connectivity improves PDR, and reduces both E2ED and AHC by effectively removing the candidate node with a weak link. The developed TDMP is suitable in supporting non-delay-tolerant applications in both highway and urban environments with heterogenous traffic and complex road layouts, such as information sharing, congestion avoidance and emergency notification.

However, current V2X communication technologies may not achieve high data rate and ultra-low delay. With the introduction of 5G technology, a mixture of multi-radio access mechanism may improve the feasibility and availability of VANET applications [60]. For extending the scale of VANET scenario, the centralised/distributed software-defined network (SDN) with the features of programmability and flexibility can be a good option. Based on the existing framework of TDMP, other traffic models could be investigated to further increase accuracy.

**References**

[1] S. Al-Sultan, M. M. Al-Doori, A. H. Al-Bayatti, and H. Zedan, “A comprehensive survey on vehicular Ad Hoc network,” *J. Netw. Comput. Appl.*, vol. 37, no. 1, pp. 380–392, 2014, doi: 10.1016/j.jnca.2013.02.036.

[2] S. Khan, M. Alam, M. Fränzle, N. Müllner, and Y. Chen, “A Traffic Aware Segment-based Routing protocol for VANETs in urban scenarios,” *Comput. Electr. Eng.*, vol. 68, pp. 447–462, May 2018, doi: 10.1016/j.compeleceng.2018.04.017.

[3] S. Boussoufa-Lahlah, F. Semchedine, and L. Bouallouche-Medjkoune, “Geographic routing protocols for Vehicular Ad hoc NETworks (VANETs): A survey,” *Veh. Commun.*, vol. 11, pp. 20–31, 2018, doi: 10.1016/j.vehcom.2018.01.006.

[4] Cisco public, “Cisco Annual Internet Report (2018–2023),” TX, USA, 2019.

[5] N. Alsharif and X. Shen, “ICAR-II: Infrastructure-based connectivity aware routing in vehicular networks,” *IEEE Trans. Veh. Technol.*, vol. 66, no. 5, pp. 4231–4244, 2017, doi: 10.1109/TVT.2016.2600481.

[6] C. Sommer, D. Eckhoff, R. German, and F. Dressler, “A computationally inexpensive empirical model of IEEE 802.11p radio shadowing in urban environments,” in *2011 8th International Conference on Wireless On-Demand Network Systems and Services, WONS 2011*, 2011, pp. 84–90, doi: 10.1109/WONS.2011.5720204.

[7] S. Sharma and A. Kaul, “A survey on Intrusion Detection Systems and Honeypot based proactive security mechanisms in VANETs and VANET Cloud,” *Vehicular Communications*, vol. 12. Elsevier Inc., pp. 138–164, 01-Apr-2018, doi: 10.1016/j.vehcom.2018.04.005.

[8] T. Nebbou, M. Lehsaini, and H. Fouchal, “Partial backwards routing protocol for VANETs,” *Veh. Commun.*, vol. 18, Aug. 2019, doi: 10.1016/j.vehcom.2019.100162.

[9] Y. Yao *et al.*, “Multi-Channel Based Sybil Attack Detection in Vehicular Ad Hoc Networks Using RSSI,” *IEEE Trans. Mob. Comput.*, vol. 18, no. 2, pp. 362–375, Feb. 2019, doi: 10.1109/TMC.2018.2833849.

[10] A. Elbery, H. S. Hassanein, N. Zorba, and H. A. Rakha, “VANET-Based Smart Navigation for Vehicle Crowds: FIFA World Cup 2022 Case Study,” in *2019 IEEE Global Communications Conference (GLOBECOM)*, 2019, pp. 1–6, doi: 10.1109/GLOBECOM38437.2019.9014183.

[11] S. Singh and S. Agrawal, “VANET routing protocols: Issues and challenges,” *2014 Recent Adv. Eng. Comput. Sci. RAECS 2014*, pp. 6–8, 2014, doi: 10.1109/RAECS.2014.6799625.

[12] B. T. Sharef, R. A. Alsaqour, and M. Ismail, “Vehicular communication ad hoc routing protocols: A survey,” *Journal of Network and Computer Applications*, vol. 40, no. 1. Academic Press, pp. 363–396, 01-Apr-2014, doi: 10.1016/j.jnca.2013.09.008.

[13] H. Hasrouny, A. E. Samhat, C. Bassil, and A. Laouiti, "VANet security challenges and solutions: A survey," *Vehicular Communications*, vol. 7. Elsevier Inc., pp. 7–20, 01-Jan-2017, doi: 10.1016/j.vehcom.2017.01.002.

[14] C. Chen, L. Liu, T. Qiu, J. Jiang, Q. Pei, and H. Song, "Routing With Traffic Awareness and Link Preference in Internet of Vehicles," *IEEE Trans. Intell. Transp. Syst.*, pp. 1–15, Aug. 2020, doi: 10.1109/tits.2020.3009455.

[15] A. Srivastava, A. Prakash, and R. Tripathi, "Location based routing protocols in VANET: Issues and existing solutions," *Veh. Commun.*, vol. 23, p. 100231, 2020, doi: 10.1016/j.vehcom.2020.100231.

[16] T. Clausen *et al.*, "Optimized Link State Routing Protocol (OLSR)," 2003.

[17] C. E. Perkins and E. M. Royer, "Ad-hoc on-demand distance vector routing," in *Proceedings - WMCSA'99: 2nd IEEE Workshop on Mobile Computing Systems and Applications*, 1999, pp. 90–100, doi: 10.1109/MCSA.1999.749281.

[18] D. Gupta and R. Kumar, "An Improved Genetic Based Routing Protocol for VANETs," *Conflu. Next Gener. Inf. Technol. Summit*, vol. 147001, pp. 347–353, 2014.

[19] D. B. Johnson, D. A. Maltz, and J. Broch, "DSR: The Dynamic Source Routing Protocol for Multi-Hop Wireless Ad Hoc Networks."

[20] N. Beijar, "Zone Routing Protocol (ZRP)."

[21] J. Liu, J. Wan, Q. Wang, P. Deng, K. Zhou, and Y. Qiao, "A survey on position-based routing for vehicular ad hoc networks," *Telecommun. Syst.*, vol. 62, no. 1, pp. 15–30, 2016, doi: 10.1007/s11235-015-9979-7.

[22] S. M. Bilal, C. J. Bernardos, and C. Guerrero, "Position-based routing in vehicular networks: A survey," *J. Netw. Comput. Appl.*, vol. 36, no. 2, pp. 685–697, 2013, doi: 10.1016/j.jnca.2012.12.023.

[23] O. S. Oubbati, A. Lakas, F. Zhou, M. Güneş, and M. B. Yagoubi, "A survey on position-based routing protocols for Flying Ad hoc Networks (FANETs)," *Vehicular Communications*, vol. 10. Elsevier Inc., pp. 29–56, 01-Oct-2017, doi: 10.1016/j.vehcom.2017.10.003.

[24] A. Ullah, X. Yao, S. Shaheen, and H. Ning, "Advances in Position Based Routing Towards ITS Enabled FoG-Oriented VANET-A Survey," *IEEE Trans. Intell. Transp. Syst.*, pp. 1–13, Feb. 2019, doi: 10.1109/tits.2019.2893067.

[25] G. Chen, C. Li, M. Ye, and J. Wu, "An unequal cluster-based routing protocol in wireless sensor networks," *Wirel. Networks*, vol. 15, no. 2, pp. 193–207, Feb. 2009, doi: 10.1007/s11276-007-0035-8.

[26] D. Lin, J. Kang, A. Squicciarini, Y. Wu, S. Gurung, and O. Tonguz, "MoZo: A Moving Zone Based Routing Protocol Using Pure V2V Communication in VANETs," *IEEE Trans. Mob. Comput.*, vol. 16, no. 5, pp. 1357–1370, May 2017, doi: 10.1109/TMC.2016.2592915.

[27] S. Ucar, S. C. Ergen, and O. Ozkasap, “VMaSC: Vehicular multi-hop algorithm for stable clustering in Vehicular Ad Hoc Networks,” in *IEEE Wireless Communications and Networking Conference, WCNC*, 2013, pp. 2381–2386, doi: 10.1109/WCNC.2013.6554933.

[28] X. Ji, H. Yu, G. Fan, H. Sun, and L. Chen, “Efficient and Reliable Cluster-Based Data Transmission for Vehicular Ad Hoc Networks,” 2018, doi: 10.1155/2018/9826782.

[29] M. Nekovee and B. B. Bogason, “Reliable and efficient information dissemination in intermittently connected vehicular adhoc networks,” in *IEEE Vehicular Technology Conference*, 2007, pp. 2486–2490, doi: 10.1109/VETECS.2007.512.

[30] A. Bachir and A. Benslimane, “A multicast protocol in ad hoc networks inter-vehicle geocast,” in *IEEE Vehicular Technology Conference*, 2003, vol. 57, no. 4, pp. 2456–2460, doi: 10.1109/vetecs.2003.1208832.

[31] C. Maihöfer, T. Leinmüller, and E. Schoch, “Abiding geocast,” in *Proceedings of the 2nd ACM international workshop on Vehicular ad hoc networks - VANET ’05*, 2005, p. 20, doi: 10.1145/1080754.1080758.

[32] A. Ullah, X. Yao, S. Shaheen, and H. Ning, “Advances in Position Based Routing Towards ITS Enabled FoG-Oriented VANET-A Survey,” *IEEE Trans. Intell. Transp. Syst.*, vol. 21, no. 2, pp. 1–13, 2019, doi: 10.1109/tits.2019.2893067.

[33] B. Karp and H. T. Kung, “GPSR,” in *Proceedings of the 6th annual international conference on Mobile computing and networking - MobiCom ’00*, 2000, pp. 243–254, doi: 10.1145/345910.345953.

[34] C. Lochert, H. Hartenstein, J. Tian, H. Fübler, D. Hermann, and M. Mauve, “A routing strategy for vehicular ad hoc networks in city environments,” in *IEEE Intelligent Vehicles Symposium, Proceedings*, 2003, pp. 156–161, doi: 10.1109/IVS.2003.1212901.

[35] B. C. Seet, G. Liu, B. S. Lee, C. H. Fob, K. J. Wong, and K. K. Lee, “A-STAR: A mobile Ad Hoc routing strategy for metropolis vehicular communications,” *Lect. Notes Comput. Sci. (including Subser. Lect. Notes Artif. Intell. Lect. Notes Bioinformatics)*, vol. 3042, pp. 989–999, 2004, doi: 10.1007/978-3-540-24693-0_81.

[36] C. Lochert, M. Mauve, H. Füßler, and H. Hartenstein, “Geographic routing in city scenarios,” *ACM SIGMOBILE Mob. Comput. Commun. Rev.*, vol. 9, no. 1, p. 69, Jan. 2005, doi: 10.1145/1055959.1055970.

[37] M. Jerbi, S.-M. Senouci, M. Jerbi, R. Meraihi, and Y. Ghamri-Doudane, “GyTAR: improved greedy traffic aware routing protocol for vehicular ad hoc networks in city environments Characterization of a delay and disruption tolerant network in the Amazon basin View project Vehicular networks/Internet of Vehicles/Vehicular Cloud n,” 2006, doi: 10.1145/1161064.1161080.

[38] Y. Feng, F. Wang, J. Liao, and Q. Qian, "Driving Path Predication Based Routing Protocol in Vehicular Ad hoc Networks," *Int. J. Distrib. Sens. Networks*, vol. 2013, 2013, doi: 10.1155/2013/837381.

[39] T. Lu, S. Chang, and W. Li, "Fog computing enabling geographic routing for urban area vehicular network," *Peer-to-Peer Netw. Appl.*, vol. 11, no. 4, pp. 749–755, Jul. 2018, doi: 10.1007/s12083-017-0560-x.

[40] X. Wang, C. Li, L. Zhu, and C. Zhao, "An effective routing protocol for intermittently connected vehicular ad hoc networks," in *IEEE Wireless Communications and Networking Conference, WCNC*, 2013, pp. 1750–1755, doi: 10.1109/WCNC.2013.6554828.

[41] N. Goel, I. Dhyani, and G. Sharma, "An acute position based VANET routing protocol," in *Proceedings - 2016 International Conference on Micro-Electronics and Telecommunication Engineering, ICMETE 2016*, 2016, pp. 139–144, doi: 10.1109/ICMETE.2016.109.

[42] A. Guleria and K. Singh, "Position based adaptive routing for VANETs," *Artic. Int. J. Comput. Networks Commun.*, vol. 9, no. 1, 2017, doi: 10.5121/ijcnc.2017.9105.

[43] K. N. Qureshi, A. H. Abdullah, and J. Lloret, "Road Perception Based Geographical Routing Protocol for Vehicular Ad Hoc Networks:," *http://dx.doi.org/10.1155/2016/2617480*, Feb. 2016, doi: 10.1155/2016/2617480.

[44] M. Sadou and L. Bouallouche-Medjkoune, "Efficient message delivery in hybrid sensor and vehicular networks based on mathematical linear programming," *Comput. Electr. Eng.*, vol. 64, pp. 496–505, Nov. 2017, doi: 10.1016/j.compeleceng.2016.11.032.

[45] H. Saleet, R. Langar, K. Naik, R. Boutaba, A. Nayak, and N. Goel, "Intersection-based geographical routing protocol for VANETs: A proposal and analysis," *IEEE Trans. Veh. Technol.*, vol. 60, no. 9, pp. 4560–4574, 2011, doi: 10.1109/TVT.2011.2173510.

[46] J. Gong, C. Z. Xu, and J. Holle, "Predictive directional greedy routing in vehicular ad hoc networks," in *Proceedings - International Conference on Distributed Computing Systems*, 2007, doi: 10.1109/ICDCSW.2007.65.

[47] R. Karimi and S. Shokrollahi, "PGRP: Predictive geographic routing protocol for VANETs," *Comput. Networks*, vol. 141, pp. 67–81, 2018, doi: 10.1016/j.comnet.2018.05.017.

[48] M. Ye, L. Guan, and M. Quddus, "MPBRP-mobility prediction based routing protocol in VANETs," in *Proceedings - 2019 International Conference on Advanced Communication Technologies and Networking, CommNet 2019*, 2019, doi: 10.1109/COMMNET.2019.8742389.

[49] J. Zhao and G. Cao, "VADD: Vehicle-assisted data delivery in vehicular Ad hoc networks," *IEEE Trans. Veh. Technol.*, vol. 57, no. 3, pp. 1910–1922, May 2008, doi: 10.1109/TVT.2007.901869.

[50] I. Leontiadis and C. Mascolo, "GeOpps: Geographical opportunistic routing for vehicular networks," in *2007 IEEE International Symposium on a World of Wireless, Mobile and Multimedia Networks, WOWMOM*, 2007, doi: 10.1109/WOWMOM.2007.4351688.

[51] M. Jerbi, S. M. Senouci, T. Rasheed, and Y. Ghamri-Doudane, "Towards efficient geographic routing in urban vehicular networks," *IEEE Trans. Veh. Technol.*, vol. 58, no. 9, pp. 5048–5059, 2009, doi: 10.1109/TVT.2009.2024341.

[52] J. Liu *et al.*, "High-Efficiency Urban Traffic Management in Context-Aware Computing and 5G Communication," *IEEE Commun. Mag.*, vol. 55, no. 1, pp. 34–40, Jan. 2017, doi: 10.1109/MCOM.2017.1600371CM.

[53] F. Goudarzi, "Non-Cooperative Beaconing Control in Vehicular Ad hoc Networks," Brunel University London, 2017.

[54] C. Sommer *et al.*, "Veins: The Open Source Vehicular Network Simulation Framework," Springer, Cham, 2019, pp. 215–252.

[55] C. Sommer, R. German, and F. Dressler, "Bidirectionally coupled network and road simulation for improved IVC analysis," *IEEE Trans. Mob. Comput.*, vol. 10, no. 1, pp. 3–15, Jan. 2011, doi: 10.1109/TMC.2010.133.

[56] P. A. Lopez *et al.*, "Microscopic Traffic Simulation using SUMO," in *IEEE Conference on Intelligent Transportation Systems, Proceedings, ITSC*, 2018, vol. 2018-Novem, pp. 2575–2582, doi: 10.1109/ITSC.2018.8569938.

[57] A. Varga and R. Hornig, "An overview of the OMNeT++ simulation environment," doi: 10.1145/1416222.1416290.

[58] J. Song, Y. Wu, Z. Xu, and X. Lin, "Research on car-following model based on SUMO," in *Proceedings of 2014 IEEE 7th International Conference on Advanced Infocomm Technology, IEEE/ICAIT 2014*, 2015, pp. 47–55, doi: 10.1109/ICAIT.2014.7019528.

[59] M. Treiber and V. Kanagaraj, "Comparing numerical integration schemes for time-continuous car-following models," *Phys. A Stat. Mech. its Appl.*, vol. 419, pp. 183–195, Feb. 2015, doi: 10.1016/j.physa.2014.09.061.

[60] P. K. Singh, S. K. Nandi, and S. Nandi, "A tutorial survey on vehicular communication state of the art, and future research directions," *Veh. Commun.*, vol. 18, p. 100164, Aug. 2019, doi: 10.1016/j.vehcom.2019.100164.